\documentclass[aps,prc,twocolumn,showpacs,superscriptaddress,nofootinbib,longbibliography,preprintnumbers]{revtex4-1}
\usepackage[utf8]{inputenc}

\usepackage{amsmath}
\usepackage{amssymb}
\usepackage{inputenc}
\usepackage{amssymb}
\usepackage{graphicx}
\usepackage{epsfig}
\usepackage{bm}
\usepackage{color}
\usepackage{float}
\usepackage{dcolumn}
\usepackage{multirow} 
\usepackage{MnSymbol}
\usepackage[unicode=true,
  linktocpage,
  linkbordercolor={0.5 0.5 1},
  citebordercolor={0.5 1 0.5},
  colorlinks=true,
  linkcolor=blue,
  citecolor=blue,
  urlcolor=blue]{hyperref} 
\usepackage{lipsum} 
\usepackage{placeins} 
\usepackage{mathtools}

\usepackage[dvipsnames,usenames]{xcolor}

\begin{document}

\preprint{IQuS@UW-21-010}

\title{Spectral density reconstruction with Chebyshev polynomials}
\date{\today}

\author{Joanna~E.~Sobczyk}
\email{jsobczyk@uni-mainz.de}
\affiliation{Institut f\"ur Kernphysik and PRISMA$^+$ Cluster of Excellence, Johannes Gutenberg-Universit\"at, 55128
  Mainz, Germany}

\author{Alessandro Roggero}
\email{a.roggerog@unitn.it}
\affiliation{Physics Department, University of Trento, Via Sommarive 14, I-38123 Trento, Italy}
\affiliation{INFN-TIFPA Trento Institute of Fundamental Physics and Applications,  Trento, Italy}
\affiliation{InQubator for Quantum Simulation (IQuS), Department of Physics, University of Washington, Seattle, WA 98195, USA}

\begin{abstract}
Accurate calculations of the spectral density in a strongly correlated quantum many-body system are of fundamental importance to study its dynamics in the linear response regime. Typical examples are the calculation of inclusive and semi-exclusive scattering cross sections in atomic nuclei and transport properties of nuclear and neutron star matter. 
Integral transform techniques play an important role in accessing the spectral density in a variety of nuclear systems. However, their accuracy is in practice limited by the need to perform a numerical inversion which is often ill-conditioned. 
In the present work we extend a recently proposed quantum algorithm which circumvents this problem.  We show how to perform controllable reconstructions of the spectral density over a finite energy resolution with rigorous error estimates. An appropriate expansion in Chebyshev polynomials allows for efficient simulations also on classical computers.
We apply our idea to reconstruct a simple model -- response function as a proof of principle. This paves the way for future applications in nuclear and condensed matter physics.

\end{abstract}

\maketitle

\section{Introduction}

A major challenge in nuclear many-body theory is the accurate prediction of scattering cross sections in low energy reactions involving both atomic nuclei and infinite nuclear matter. For ab-initio approaches with a strong connection to the underlying theory of QCD it is fundamental to be able to control the approximation errors in both the employed interactions and the adopted many-body method. With the help of an effective field theory approach the first of this sources of uncertainty has started to be put on a firmer ground~\cite{Weinberg:1990rz,Weinberg:1991um,Ordonez:1995rz,Kaplan:1996xu,Epelbaum:1998hg,Machleidt_2011} and theoretical error estimates coming from the modelling of nuclear interactions are now an integral part of the work of nuclear theorists~\cite{McDonnell2015,Wesolowski_2016,Melendez2017,Drischler2020}. Using similar tools great efforts are being pursued by the nuclear theory community to understand the systematic errors introduced by the approximate many-body techniques used to solve the nuclear ground states~\cite{Ekstrom2019,KONIG2020}. Benchmark calculations for ground state properties of few-body nuclei have also been performed (see e.g.~\cite{Kamada2001}) but a more complete understanding of the various sources of systematic errors in predictions of nuclear dynamics for larger systems is hindered by the incredible computational complexity of the problem.

A very powerful approach to study dynamical properties in medium-mass systems and infinite matter is the adoption of integral transform techniques which map the local density of states into more manageable integrated quantities, a typical example being sum-rules of the nuclear response which describe moments of the density of states and can be expressed directly as ground-state expectation values~\cite{Orlandini_1991,Sobczyk2020}. This important map between real-time observables and ground-state expectation values can be achieved more generally by employing integral transforms with various kernels followed by a numerical inversion of the resulting integral transform to recover the response function in the frequency domain. The choice of integral kernel is typically dictated by the possibility of evaluating the ensuing integral transform with a powerful many-body technique. Two of the most popular examples are the Lorentz Integral Transform (LIT) widely used in conjunction with diagonalization techniques~\cite{Efros_1994,Efros_2007} and, more recently, with the coupled cluster method  (LIT-CC)~\cite{Bacca2013,Bacca2014,sobczyk2021ab} and the Laplace transform applied with Monte Carlo methods thanks to its relationship with imaginary-time correlation functions~\cite{Carlson_1992,Carlson2015,Lovato_2016}. A crucial component of these approaches consists in inverting the integral transform, a process that for the Laplace transform can be seen as analytical continuation from imaginary-time to the real time axis~\cite{Jarrell1996}. In general this procedure when applied to invert an integral transform obtained by numerical methods is ill-posed, in the sense that small errors in the input response can give rise to arbitrarily large high-frequency noise in the reconstructed real frequency response~\cite{Gl_ckle_2009,Barnea_2010}. A variety of approximate inversion techniques that introduce, more or less explicitly, additional smoothing to reduce these high-frequency oscillations have been proposed in the past~\cite{Silver1990,Vitali2010,Burnier2013,Kades2020,Raghavan2021}. These can be very successful in situations where the dominant structure of the response function is simple and known beforehand, such as for the quasi-elastic peak in medium energy scattering~\cite{Lovato_2016,Lovato2020}, but the introduced systematic errors are no longer sufficiently under control to trust predictions with unexpected features thus severely limiting explorations of the nuclear dynamics in challenging regimes where little experimental information is available to guide the inversion.

At this point it is important to mention that some observables connected with integrated properties of the response, like e.g. the electric dipole polarizability of nuclei~\cite{Miorelli2016} or the impurity contribution to the thermal conductivity in the outer crust of neutron stars~\cite{Roggero_2016}, can be obtained directly from the integral transform thus allowing to avoid the inversion step. Moreover, the severity of the induced systematic errors strongly depends on the properties of the integral kernel that defines the integral transform, a feature recognized early on and one of the inspirations for introducing the LIT in nuclear physics~\cite{Efros_1994,Leidemann_2015} as well as generalizations of the Laplace transform~\cite{Roggero_2013,Rota_2015}. One of the salient features of an ideal integral transform kernel is the ability to set a resolution scale which then allows for an effective coarse-graining of the frequency space signal, e.g. for the LIT this is controlled by the kernel width. This intuition led recently to the introduction of quantum algorithms to reliably estimate both inclusive and exclusive scattering cross sections through an appropriate integral transform of the spectral density using simulations performed with quantum computers~\cite{roggero2019,roggero2020C,roggero2020A} (see also~\cite{Choi2021,Somma_2019,Rall2020} for similar approaches and~\cite{Ba_uls_2020,klco2021standard} for recent reviews).

In this work we extend the results of Ref.~\cite{roggero2020C} to show how to reconstruct, with controllable errors, general response function in frequency space from integral transforms expressed on a basis of Chebyshev polynomials thus completely avoiding the use of uncontrollable numerical inversion procedures. The use of Chebyshev polynomials for this task is reminiscent of the Kernel Polynomial Method (KPM)~\cite{Wei_e_2006} introduced in the context of condensed matter physics and especially popular in conjunction with a matrix product representation (see e.g.~\cite{Yang_2020,Papaefstathiou_2021}). 

The paper is organized as follows. In the next Section we briefly describe the approach introduced in Ref.~\cite{roggero2020C} for the calculation of the spectral density discussing differences and similarities with KPM. The method's accuracy and its dependence on the particular choice of integral kernel is discussed in Sec.~\ref{sec:pwise_convergence} where we also compare it directly with the more standard KPM approach. In Sec.~\ref{sec:histo} we introduce a new construction for coarse-graining the spectral density in a way that allows for a direct control of the approximation error and study a simple benchmark to show its efficacy. Finally, in Sec.~\ref{sec:summary} we conclude and discuss the potential benefit of our proposal when used in conjunction with classical many-body techniques like matrix product states and coupled cluster theory.

\section{Formalism}

Following the presentation in Ref.~\cite{roggero2020C} we start by introducing the local density of states (or dynamical response function) defined as
\begin{equation}
S(\omega) = \frac{\langle\Psi_0\lvert \hat{O}\delta\left(\hat{H}-\omega\right)\hat{O}\rvert\Psi_0\rangle}{\langle\Psi_0\lvert\hat{O}^2\rvert\Psi_0\rangle}\;, 
\end{equation}
where $\rvert\Psi_0\rangle$ is the ground-state, $\hat{O}$ is an (hermitian) excitation operator describing the scattering vertex and $\hat{H}$ is the nuclear Hamiltonian. Note that with this definition the density of states is normalized as $\int d\omega S(\omega) = 1$. For finite systems the Hamiltonian spectrum is discrete by construction but here we consider $\omega$ as a continuous variable by employing the Dirac delta function as follows
\begin{equation}
\label{eq:spectral_density_expansion}
S(\omega) =\! \sum_n\! \frac{\rvert\langle \Psi_0\lvert\hat{O}\rvert \phi_n\rangle\lvert^2}{\langle\Psi_0\lvert\hat{O}^2\rvert\Psi_0\rangle} \delta\left(E_n-\omega\right)=\sum_n s_n \delta\left(E_n-\omega\right),
\end{equation}
with $\rvert\phi_n\rangle$ energy eigenstates with eigenvalues $E_n$. Furthermore, we will assume that the Hamiltonian has been normalized so that the entire spectrum $\{E_n\}$ is contained in the interval $[-1,1]$. As explained in the introduction, the main focus of this work will be an integral transform $\Phi(\nu)$ of the response function defined through an integral kernel $K(\nu,\omega)$ as
\begin{equation}
\label{eq:int_transf}
\Phi(\nu) = \int_{-\infty}^\infty d\omega K(\nu,\omega) S(\omega)\;.
\end{equation}
In this work we will focus on translationally invariant integral kernels that depend only on the absolute value of the energy difference $K(\nu,\omega)\equiv K(|\omega-\nu|)$ but the results described here can be easily extended to the general case.
For ease of derivation the limits of integration extend to $\pm \infty$, with the understanding that $S(\omega)=0$ for $\lvert\omega\rvert>1$. In order to simplify the notation we will avoid to specify these limits when there is no ambiguity. We are in general interested in observables that can be expressed as energy integrals of the local density of state $S(\omega)$ as
\begin{equation}
\label{eq:observ}
Q(S,f) = \int d\omega S(\omega)f(\omega)
\end{equation}
with a bounded function $f(\omega)$ defining the specific observable under consideration. Using the integral transform $\Phi$ introduced in Eq.~\eqref{eq:int_transf} we can define the quantity
\begin{equation}
\begin{split}
\label{eq:observ}
Q(\Phi,f) &= \int d\nu \Phi(\nu)f(\nu)\\
&= \int d\nu \int d\omega K(|\omega-\nu|)S(\omega)f(\nu)\\
&=  \int d\omega \left(\int d\nu K(|\omega-\nu|)f(\nu)\right)S(\omega)\\
&=  \int d\omega \widetilde{f}(\omega)S(\omega)=Q(S,\widetilde{f})\;.
\end{split}
\end{equation}
Our goal is to determine the conditions for which the latter is a good approximation to the original observable
\begin{equation}
\left\lvert Q(\Phi,f)-Q(S,f)\right\rvert=\left\lvert Q(S,\widetilde{f})-Q(S,f)\right\rvert \leq \epsilon\;,
\end{equation}
with bounded error $\epsilon>0$. For this purpose, it is convenient to define integral kernels to be $\Sigma$-\textit{accurate} with \textit{resolution} $\Lambda$ (see also Ref.~\cite{roggero2020C}) if the following holds
\begin{equation}
\inf_{\omega_0\in[-1,1]}\int_{\omega_0-\Lambda}^{\omega_0+\Lambda} d\nu K(\nu, \omega_0) \geq 1-\Sigma\;.
\label{eq:kernel_definition}
\end{equation}
As shown in Ref.~\cite{roggero2020C}, for this class of kernels we have
\begin{equation}
\epsilon \leq f^\Lambda_{max} + 2\Sigma \sup_{\omega\in[-1,1]}\left|f(\omega)\right|\;,
\end{equation}
with $f^\Lambda_{max}$ the modulus of continuity given by
\begin{equation}
f^{\Lambda}_{max} = \sup_{\omega\in[-1,1]} \sup_{x\in[-\Lambda,\Lambda]} \left|f(\omega+x)-f(\omega)\right|\;.
\end{equation}

In this work we will consider two types of translationally invariant integral kernels with an energy resolution controlled by an external parameter $\lambda$:
\begin{itemize}
    \item the Lorentzian kernel describing the Lorentz Integral Transform (LIT) from Ref.~\cite{Efros_1994}
\begin{equation}
\label{eq:lorentzianKer}
    K^{(L)}(\nu,\omega; \lambda) = \frac{1}{\pi \lambda}\frac{\lambda^2}{(\omega-\nu)^2+\lambda^2}\;,
\end{equation}
    \item the Gaussian kernel giving the Gaussian Integral Transform (GIT) from Ref.~\cite{roggero2020C}
\begin{equation}
\label{eq:gaussianKer}
    K^{(G)}(\nu,\omega; \lambda) = \frac{1}{\sqrt{2\pi} \lambda} \exp\left(-\frac{(\omega-\nu)^2}{2\lambda^2}\right)\;.
\end{equation}    
\end{itemize}

In order to evaluate the integral transform $\Phi(\nu)$ using a suitable many-body method, we will consider an expansion of these kernels into a complete basis of orthogonal polynomials $\{T_k(\omega)\}$ as
\begin{equation}
K(\nu,\omega;\lambda) = \sum_k^\infty c_k(\nu;\lambda) T_k(\omega)\;,
\label{eq:kernel_expansion}
\end{equation}
with real coefficients $c_k(\nu;\lambda)$ depending both on the location in energy $\nu$ and the kernel resolution $\lambda$. With this representation we can now express the integral transform as a linear combination
\begin{equation}
\begin{split}
\Phi(\nu;\lambda) &= \int d\omega K(\nu,\omega;\lambda) S(\omega)=\sum_k^\infty c_k(\nu;\lambda) m_k\\
\end{split}
\end{equation}
with generalized moments defined as
\begin{equation}
m_k = \int d\omega T_k(\omega) S(\omega) =\frac{\langle\Psi_0\lvert \hat{O}T_k\left(\hat{H}\right)\hat{O}\rvert\Psi_0\rangle}{\langle\Psi_0\lvert\hat{O}^2\rvert\Psi_0\rangle}
\label{eq:moments_def}
\end{equation}
and independent on the specific integral kernel employed in the construction. This property is particularly advantageous since, once the moments $\{m_k\}$ are computed with the many-body method of choice, it allows to consider a variety of integral transforms in post-processing.

In practice, only a limited number $N$ of moments will be available with a finite computational effort. We will then consider approximations to integral transforms obtained by a finite truncation of the series expansion
\begin{equation}
\Phi_N(\nu;\lambda) = \sum_k^N c_k(\nu;\lambda) m_k\;,
\label{eq:series_exp}
\end{equation}
leading to a finite approximation accuracy 
\begin{equation}
\label{eq:beta}
\sup_{\nu\in[-1,1]}\left\lvert\Phi(\nu;\lambda) - \Phi_N(\nu;\lambda)\right\rvert\leq\beta\;,
\end{equation}
with constant $\beta>0$. Ideal kernels, like the Gaussian, have a fast (i.e. exponential) convergence of $\beta$ with the number of terms $N$. The choice of the polynomial basis $\{T_k\}$ influences this convergence rate. In this work we use the Chebyshev polynomials of the first kind thanks to their quick convergence for smooth functions and we will refer to our method as CheET (Chebyshev Expansion of integral Transform). An explicit derivation of the coefficients $c_k(\nu;\lambda)$ for both the Lorentzian and Gaussian kernels can be found in Appendix~\ref{app_A}.

\begin{figure*}[t]
  \includegraphics[width=0.49\textwidth]{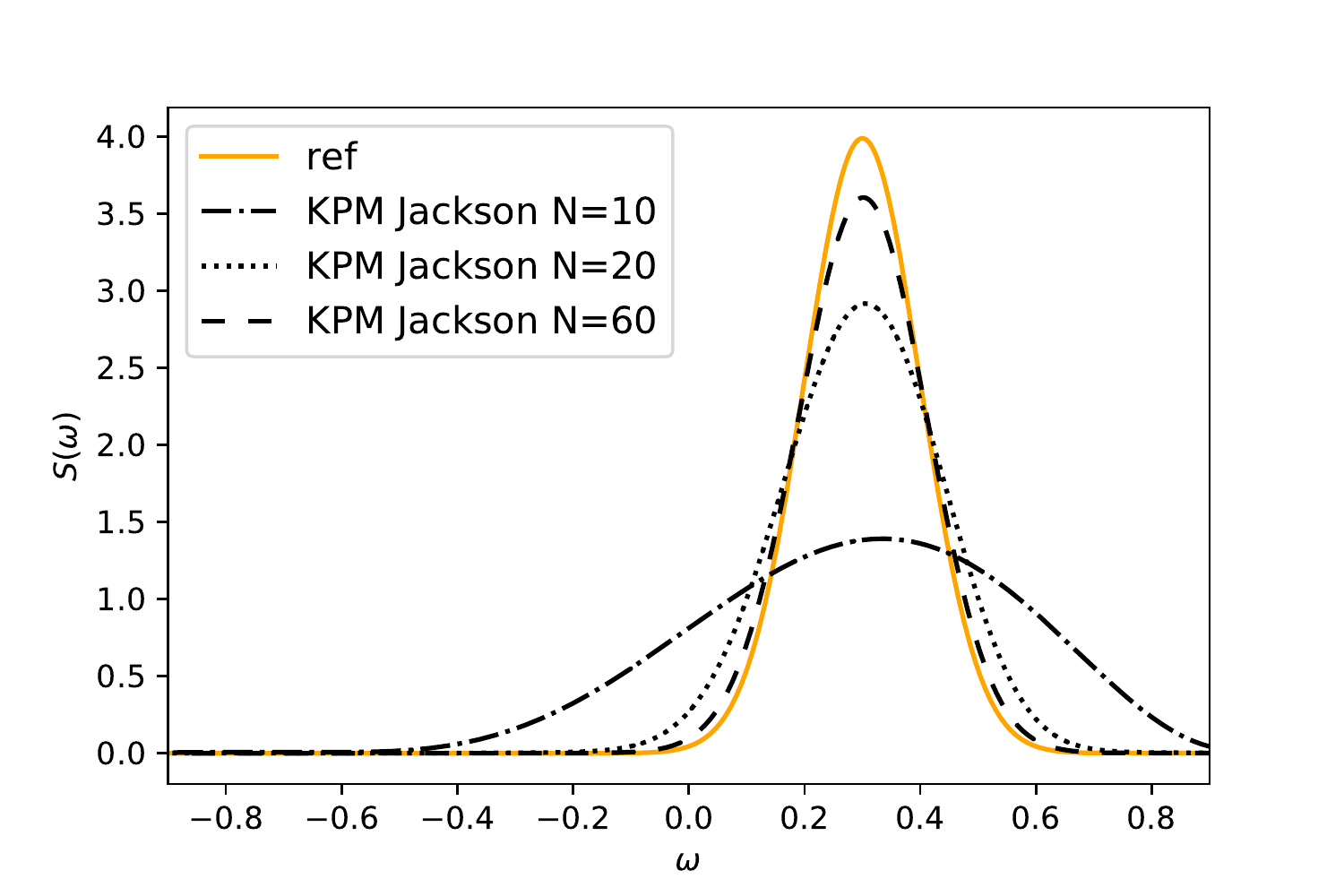}

    \includegraphics[width=0.49\textwidth]{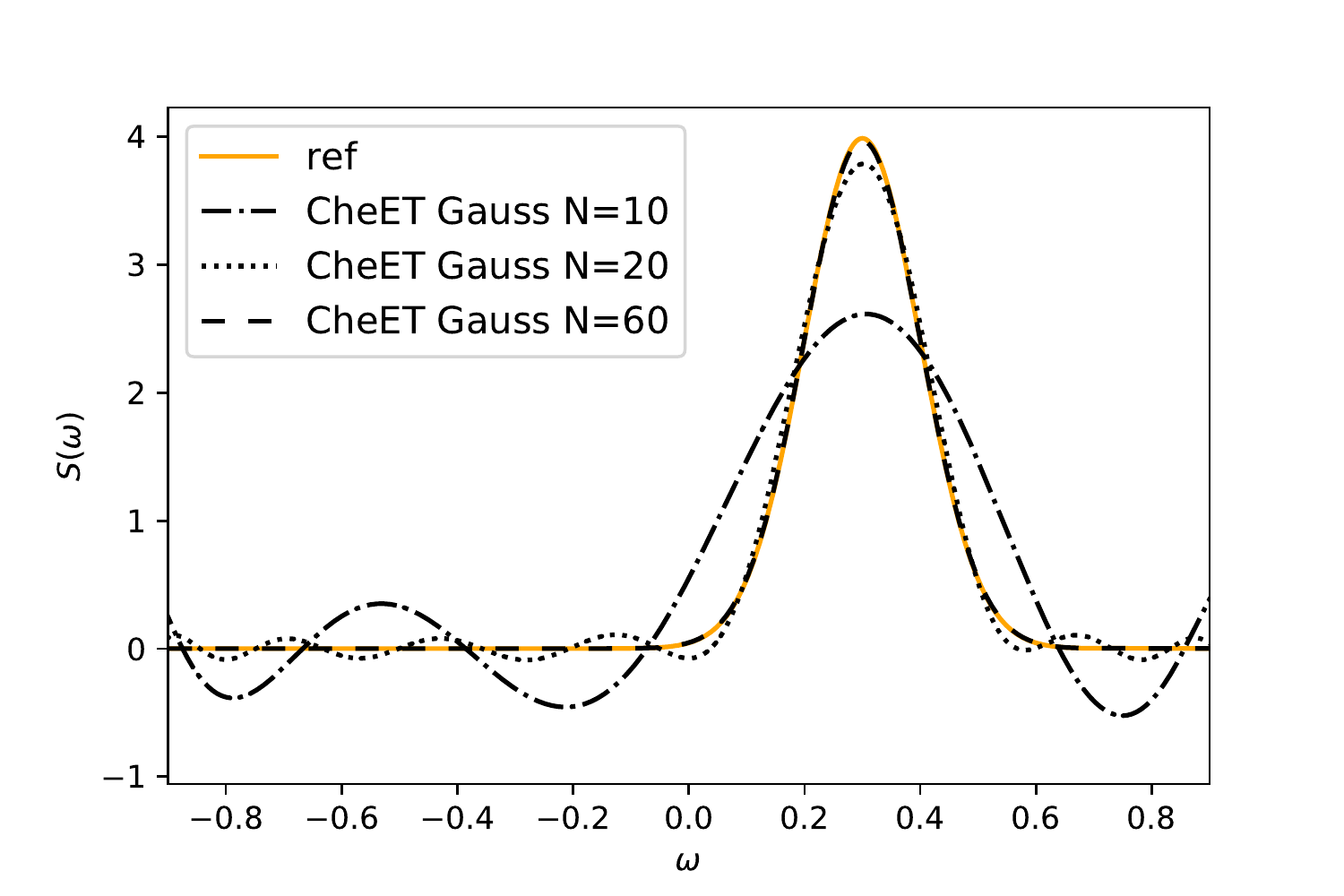}

  \caption{Reconstruction of a Gaussian signal $S(\omega)$ of a width $\Gamma=0.1$ centered at $\eta=0.3$. In the case of CheET method, $\lambda=0.01$ is being used.}
  \label{fig:gaussian-convergence}
\end{figure*}

\subsection{Evaluation of Chebyshev moments}

Chebyshev polynomials of the first kind are defined in the interval $[-1,1]$ as $T_k(\omega) = \cos [k\arccos(\omega) ]$. They follow a recursive relation
\begin{equation}
\begin{split}
 &T_0(x) = 1;\, \, \, \, \,T_{-1}(x) = T_1(x) = x;\\
 &T_{n+1}(x) = 2x T_n(x) - T_{n-1}(x)\;.
\label{eq:chebyshev}   
\end{split}    
\end{equation}
The moments of the expansion $m_k$ from Eq.~\eqref{eq:moments_def} can be retrieved using this relation:
\begin{equation}
\begin{split}
    & |\Psi_1 \rangle \equiv \hat{O} |\Psi_0\rangle\, ,\ \ \ \ \ \ |\Psi_n\rangle = \hat H |\Psi_{n-1}\rangle\\
    & m_{0} =  \langle  \Psi_1|\Psi_1 \rangle \, ,\ \ \ \ \ m_1 =   \langle \Psi_1| \Psi_2 \rangle \equiv \langle \Psi_2| \Psi_1 \rangle \\
    & m_{n+1} = 2 \langle \Psi_1| \Psi_{n+1} \rangle -m_{n-1}\equiv 2 \langle  \Psi_{n+1}| \Psi_1 \rangle-m_{n-1}\;,
\end{split}    
\end{equation}
which is particularly suited to combine with the many-body methods for which it is possible to iterate the action of the Hamiltonian, $\hat H|\Psi_n\rangle$. From the point of view of the numerical applications, a similar iteration has to be performed in the Lanczos procedure~\cite{Wei_e_2006}. Here, however, no orthogonality restoration is needed at each step. Consequently, at the $n$-th step only a single $|\Psi_{n}\rangle$ state has to be saved from the previous iterations. This makes the procedure faster and less memory-consuming.

As has been mentioned, in our considerations we assume that the Hamiltonian is normalized. In practical applications the range of the Hamiltonian spectrum can be obtained, e.g., via Lanczos algorithm and then rescaled so that $[E_{min}, E_{max}] \rightarrow [-1,1]$.

\subsection{Comparison with KPM}\label{sec:KPM}
The KPM, described in details in Ref.~\cite{Wei_e_2006}, can be understood as a specific approximation of Eq.~\eqref{eq:kernel_expansion} for which
\begin{equation}
    K_{\mathrm{KPM}}(\nu,\omega;\lambda) = \sum_{k}^\infty g_k(\lambda) T_k(\nu) T_k(\omega)\;,
\end{equation}
with the coefficients $g_k$ chosen in such a way that $\Phi(\omega) \xrightarrow[N\to \infty]{}S(\omega)$ and to reduce Gibbs oscillations.

A variety of $g_k$ were proposed in the past, designed to speed-up the convergance rate depending on the properties of the signal $S(\omega)$. Among them are the Jackson and Lorentz kernels defined as
\begin{equation}
\label{eq:kpm_expansion_coeff}
\begin{split}
    g_k^{\mathrm{Jackson}} = &\frac{1}{N+1}\left[  (N-k+1)\cos\frac{\pi k}{N+1} \right. \\
    &\left.+ \sin\frac{\pi k}{N+1}\cot\frac{\pi}{N+1} \right] \;,\\
    g_k^{\mathrm{Lorentz}}  =& \sinh \left( \kappa (1-k/N)\right) / \sinh(\kappa) \;,
\end{split}    
\end{equation}
which aim at approximating the Gaussian and Lorentzian shape of the kernel.
It is important to notice that the KPM coefficients $g_k$ do not depend on the resolution $\lambda$, while they are a function of the total number of moments $N$. (The parameter $\kappa$ in the case of $g_k^{\mathrm{Lorentz}}$ is introduced to mimic indirectly the $\lambda$ dependence.) In other words, $\lambda$ and $N$-dependencies become entangled in a non-trivial way.
Also the $\nu$-dependence of the $c_k$ coefficients is approximated by a single $k$-degree polynomial $T_k(\nu)$.
Although the KPM is a powerful tool which proved useful in many applications, it does not allow to control neither the resolution $\lambda$ with which we probe the spectrum, nor the errors  depending on the number of used moments.

%
%
%
%
%
%
%

\begin{figure*}[t]
  \includegraphics[width=0.49\textwidth]{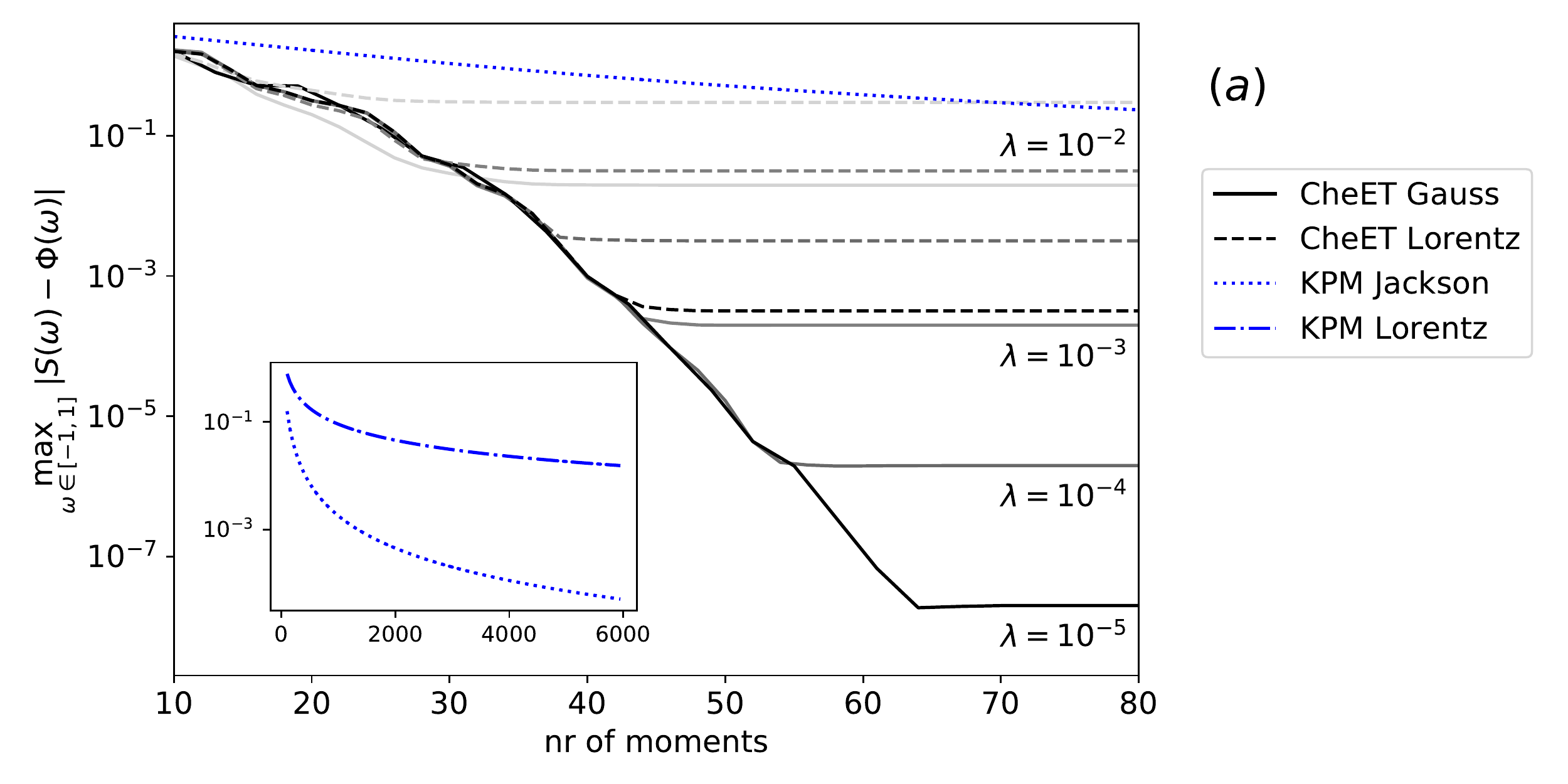}
  \includegraphics[width=0.49\textwidth]{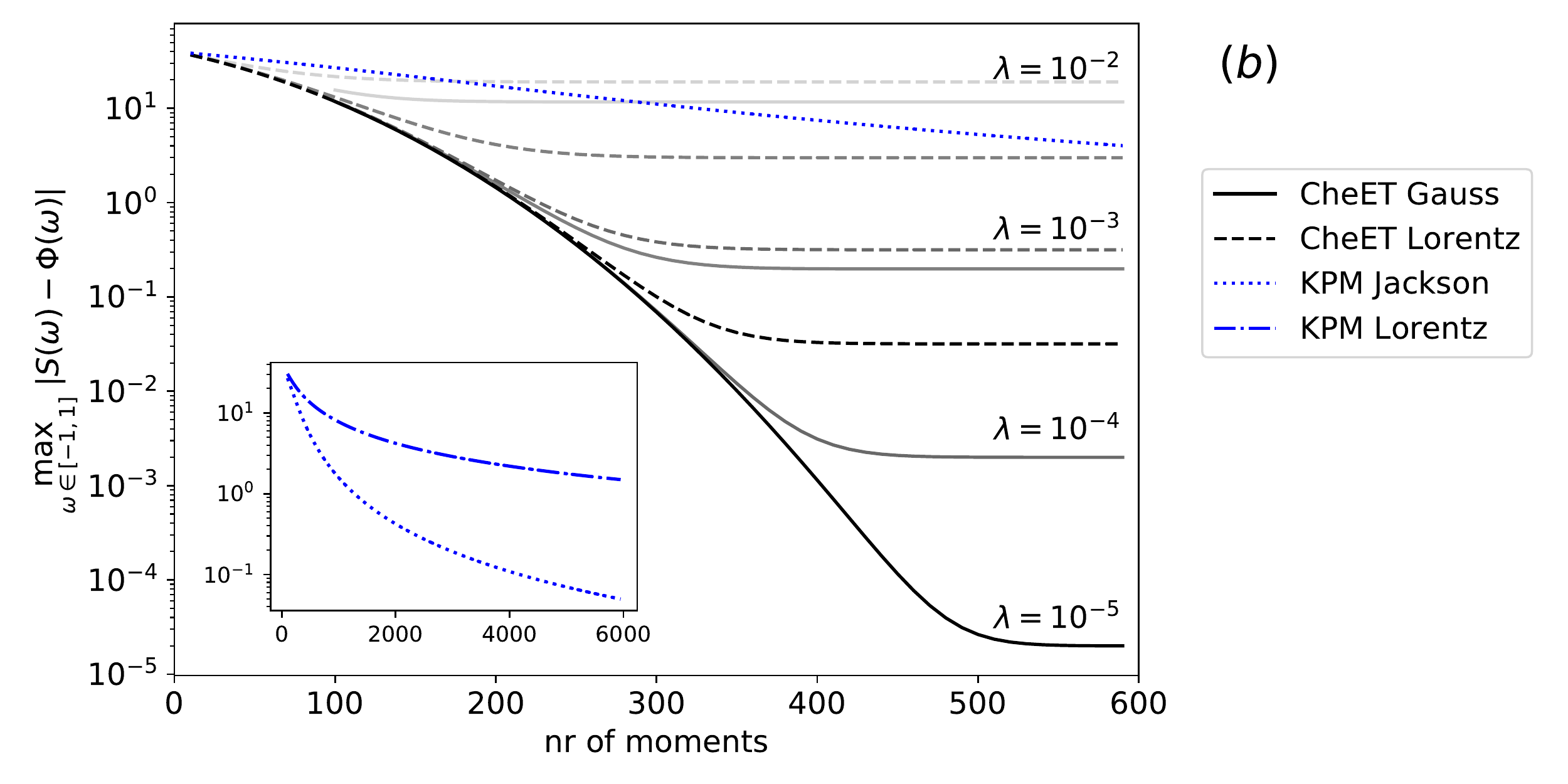}
  \caption{Comparison of the point-wise convergence of CheET and KPM methods for a Gaussian signal of width $\Gamma=0.1(0.01)$ on the left(right) panel. Solid lines correspond to CheET method with the Gaussian kernel of $\lambda$ resolution as labeled. Dashed lines correspond to the Lorentzian kernel (its resolution correspond the Gaussian). Inset shows the behaviour of KPM (both Jackson and Lorentz) for a much larger number of moments.}
  \label{fig:pointwise-methods}
\end{figure*}

\section{Analysis of point-wise convergence}
\label{sec:pwise_convergence}
To appreciate the differences between the CheET and KPM approaches, let us at first consider a continuous signal $S(\omega)$ being a single Gaussian of width $\Gamma=0.1$ or $\Gamma=0.01$ centered at $\eta=0.3$. We have checked that our conclusions hold if the signal has a Lorentzian shape or when it is composed of more than one peak (in this case the convergence pattern depends on the narrowest peak in the spectrum). Since the KPM method was primarily developed for the signal reconstruction, we will analyse the point-wise convergence. However, the main advantage of the CheET method -- the error bound -- cannot be appreciated in this case.
For the comparison to be meaningful, within the CheET we will use kernels of the width $\lambda\ll \Gamma$. 

Let us first look at the reconstruction within each of the considered methods in the case of the signal width $\Gamma=0.1$. In Fig.~\ref{fig:gaussian-convergence}, we show the results only for the Jackson and Gaussian kernels, since the behaviour of the kernels within each method (Jackson/Lorentz or Gauss/Lorentz) is qualitatively the same. The first visible distinction between the KPM  and the CheET with $\lambda=0.01$ results is the fact that the Gibbs oscillation are suppressed for the KPM approach. 
Still, the CheET Gauss is visibly better converged at much lower number of moments (already for $N=60$). 

To get a more quantitative insight into the convergence pattern,
in Fig.~\ref{fig:pointwise-methods} we show the point-wise convergence as a function of used moments (and $\lambda$ for the CheET), for the broad signal $\Gamma=0.1$ and the narrow $\Gamma=0.01$,
\begin{equation}
\label{eq:pw_error}
    \vartheta = \max_{\omega\in [-1,1]} |S(\omega)- \Phi(\omega)|
\end{equation}
where $\Phi$ is an integral transform of the signal $S$, as in Eq.~\eqref{eq:int_transf}.
We use $\kappa=3$ for the Lorentz KPM (see Eq.~\eqref{eq:kpm_expansion_coeff}).
%
The CheET curves follow a characteristic pattern: after the initial steep-slope convergence, they reach a plateau. Further addition of moments would improve the kernel reconstruction (and thus diminish the truncation error), however this does not affect the quality of signal approximation. The exact number of moments needed to reach the plateau, depends on the signal itself, in particular on the signal's resolution $\Gamma$. This can be seen when comparing both panels of Fig.~\ref{fig:pointwise-methods}. Using kernels of the same resolution $\lambda$, not only a higher number of moments is needed for smaller $\Gamma$, but also the plateau is reached with different accuracy. For $\lambda=0.001$, the CheET Gauss reaches an accuracy $\sim e^{-8}(e^{-1})$ for $\Gamma=0.1(0.01)$.
A direct comparison of CheET with the Lorentzian and Gaussian kernels with the same width $\lambda$, clearly shows that in the latter case the convergence is orders of magnitude better. In order to obtain results of a similar precision, we would need to use a Lorentzian of a much smaller width. This consequently requires a larger number of moments, if we are to control the truncation error.

In the case of the point-wise error considered here, neither KPM nor CheET are able to give a full theoretical uncertainty estimation.
Using the KPM method has an advantage that with the increasing number of moments $N$, we converge to the original signal $\Phi_{\mathrm{KPM}}^N(\omega) \xrightarrow[{N\rightarrow\infty}]{} S(\omega)$. However, the pace of convergence or the approximation error is unknown. From our observations, the CheET rate of convergence (before reaching a plateau) is much faster than KPM.
Although for a given resolution $\lambda$ the CheET method reaches a plateau, in the limit $N\to\infty$ the CheET predictions can also converge to the original signal. In order to achieve this we should progressively reduce the resolution $\lambda$ of the kernel with the increasing number of moments. We may do so by appropriately scaling $\lambda(N)$, e.g. by keeping fixed the truncation error at a satisfactory low value.  
%

For the CheET, we are still able to provide an estimate for the truncation error $\beta$  as a function of number of moments used for the reconstruction of the kernel. This can be obtained, even without knowing the value of moments $m_k$ for $k>N$, using the following bound
\begin{equation}
\begin{split}
\label{eq:beta_bound}
\left\lvert\Phi(\nu;\lambda) - \Phi_N(\nu;\lambda)\right\rvert &= \left\lvert \sum_{k=N+1}^\infty c_k(\nu;\lambda) m_k \right\rvert\\
&\leq \sum_{k=N+1}^\infty \left\lvert c_k(\nu;\lambda) \right\rvert\;,
\end{split}
\end{equation}
which, when maximized over $\nu$, gives in turn an upperbound on the truncation error $\beta$ defined in Eq.~\eqref{eq:beta}.
In Fig.~\ref{fig:beta-error}, we show the bound obtained in this way for eight values of $\lambda$ (see Appendix~\ref{app:truncation_err} for closed-form expressions for these). As expected, the Gaussian kernel performs better than the Lorentzian. For the truncation error to be at the order of $0.1$ and $\lambda=0.01$ one needs $N\approx 900$ moments. To go an order of magnitude further to $\lambda=0.001$, the number of moments increases correspondingly to $N\approx 10000$. When comparing these numbers with Fig.~\ref{fig:pointwise-methods}, we realize that the plateau is reached much faster, even below $N=100$. This discrepancy is likely coming from the use of the bound in Eq.~\eqref{eq:beta_bound} which erases structural information from the moments and therefore assumes the truncation is for a worse-case scenario signal of width $\approx\lambda$ instead.

\begin{figure}[t]
    \includegraphics[width=0.5\textwidth]{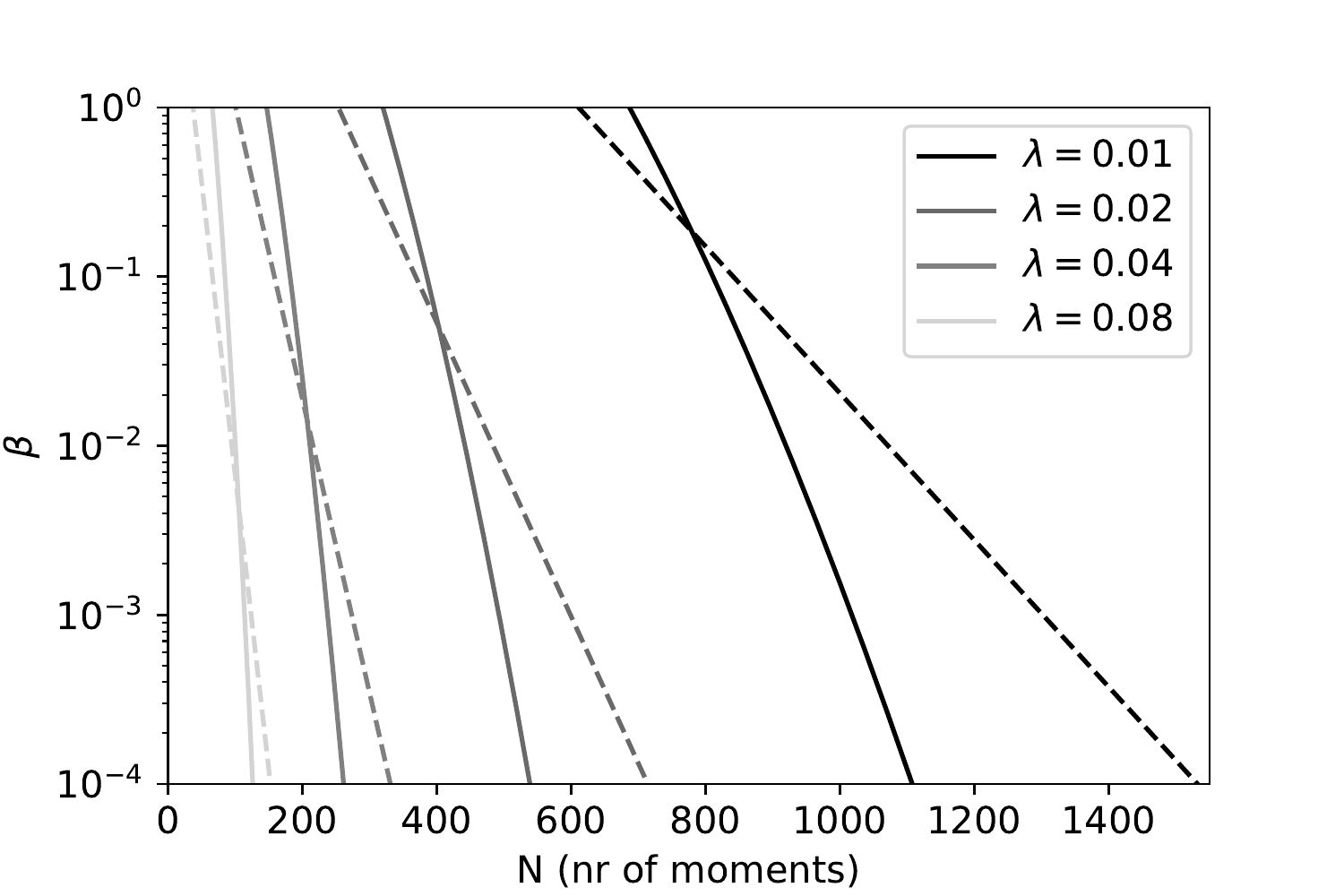}  
    \includegraphics[width=0.5\textwidth]{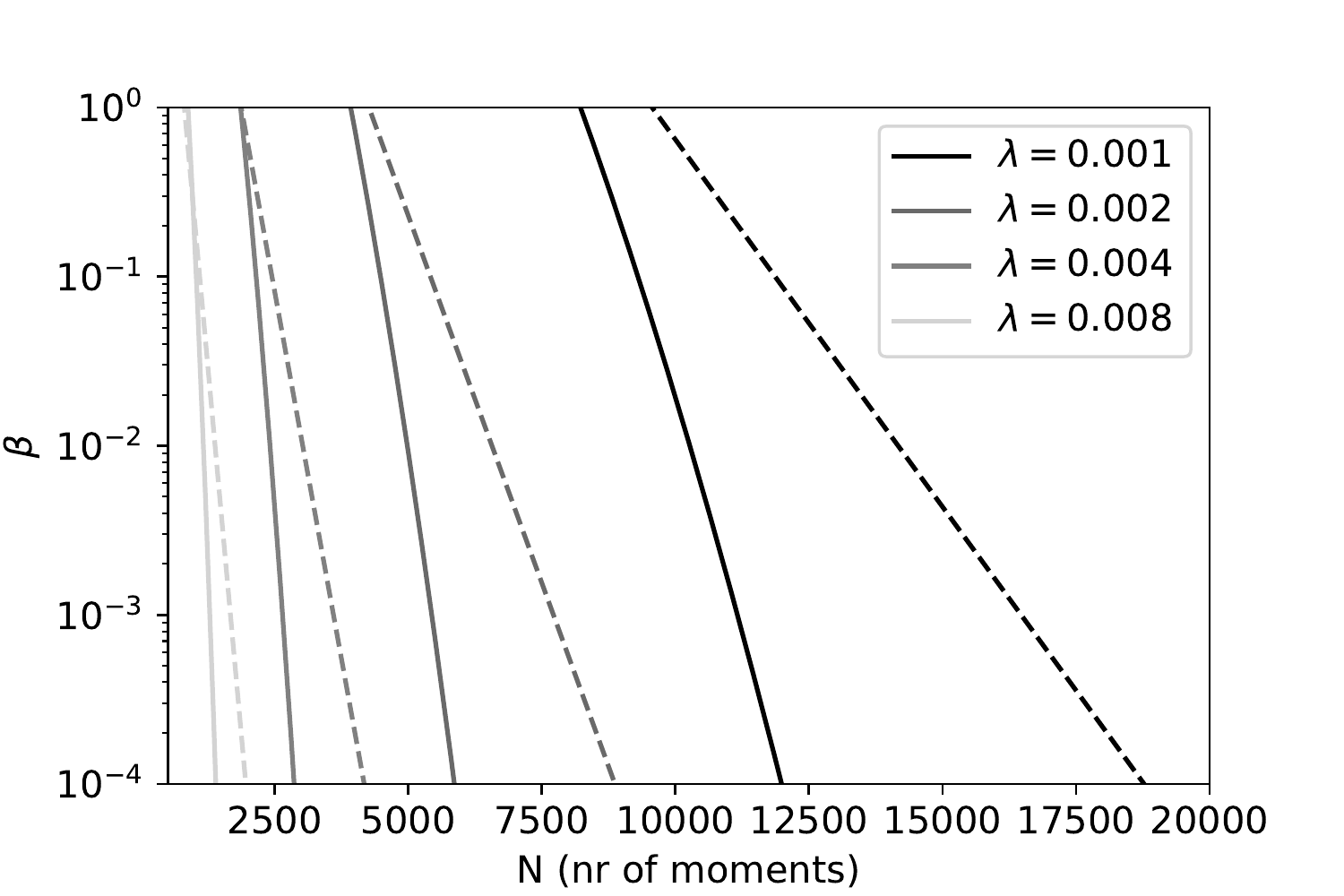} 
  \caption{Truncation error as a function of number of moments. Solid(dashed) line corresponds to CheET Gaussian(Lorentzian) for eight resolutions $\lambda$.}
  \label{fig:beta-error}
\end{figure}

Let us come back to a remark done in Sec.~\ref{sec:KPM}. While in the case of KPM (at least for the Jackson kernel), the only free parameter of the signal reconstruction is $N$, the number of moments used in the reconstruction, the CheET method introduces explicitly a smoothing scale $\lambda$ which corresponds to the regularization parameters used for the standard inversion techniques. When the signal has structures of a higher resolution, we are not able to resolve them. At first sight it might seem to be a drawback.
However, $\lambda$ gives directly the scale at which we can rely on the signal reconstruction. This is lacking in the KPM, for which in the asymptotic regime we might never see a uniform convergence of errors.
Moreover, in practical applications one has only a limited number of moments available and would like to reconstruct the signal controlling the approximation. This can be done within CheET. The resolution scale can be set depending on $N$ to keep the truncation errors sufficiently low.

\section{Direct approximation of the spectral density using histograms}
\label{sec:histo}

The results shown in the previous section are useful to gain insights into the possible benefits of using different kernel functions to study the local spectral density. For a more realistic case when the target response is not known, however, it will not be possible to compute directly the pointwise error from Eq.~\eqref{eq:pw_error} and a different, computable, error metric is needed. We achieve this by explicitly introducing a target energy resolution scale $\Delta$ and using the properties enjoyed by $\Sigma$-\textit{accurate} with \textit{resolution} $\Lambda$ integral kernels to bound the error on a suitably coarse-grained energy distribution. For this purpose, we introduce an energy histogram as a frequency observable like Eq.~\eqref{eq:observ} by defining the following window function
\begin{equation}
\label{eq:window}
f(\omega,\eta;\Delta) = \bigg\{\begin{matrix}
0 & |\eta-\omega|>\Delta\\
1 & \text{otherwise}
\end{matrix}\;.
\end{equation}
The histogram of the frequency signal $S(\omega)$, with associated bin width equal to $2\Delta$, is found by integrating over the spectrum. Explicitly, the value of the histogram centered at $\eta$ is given by
\begin{equation}
    h(\eta;\Delta) = \int_{-1}^1d\omega S(\omega)f(\omega,\eta;\Delta) = \int_{\eta-\Delta}^{\eta+\Delta}d\omega S(\omega)\;.
    \label{eq:hist_bin}
\end{equation}

We now define an approximate histogram by taking the convolution of the window function in Eq.~\eqref{eq:window} with an integral kernel with resolution $\Lambda$
\begin{equation}
\label{eq:histo_bin}
\begin{split}
\widetilde{f}^\Lambda(\omega,\eta;\Delta) &= \int_{-1}^1 d\nu K(\nu, \omega; \Lambda) f(\nu,\eta;\Delta)\\
&= \int_{\eta-\Delta}^{\eta+\Delta} d\nu K(\nu, \omega; \Lambda)\;.
\end{split}
\end{equation}
The resulting approximate histogram can be written as
\begin{equation}
\begin{split}
\widetilde{h}^\Lambda(\eta;\Delta) &= \int_{-1}^1 d\omega \widetilde{f}^\Lambda(\omega,\eta;\Delta) S(\omega)\\
&=\int_{-1}^1 d\omega  \int_{\eta-\Delta}^{\eta+\Delta} d\nu K(\nu, \omega;\Lambda) S(\omega)\;.
\end{split}
\label{eq:hist_bin_trans}
\end{equation}

Finally, we will further approximate $\widetilde{h}^\Lambda(\eta;\Delta)$ as a Chebyshev expansion truncated to order $N$ introducing an error bounded by
\begin{equation}
\label{eq:histo_truncationErr}
\sup_{\eta\in[-1,1]}\left\lvert\widetilde{h}^\Lambda(\eta;\Delta) - \widetilde{h}^\Lambda_N(\eta;\Delta)\right\rvert\leq 2\Delta \, \beta\;.
\end{equation}
Using these quantities we can then approximate the histogram with bin size $2\Delta$ at $\eta$ using the following pair of bounds
\begin{equation}
\begin{split}
&\widetilde{h}^\Lambda_N(\eta;\Delta-\Lambda) - \Sigma-2\beta\left(\Delta-\Lambda\right) \leq h(\eta,\Delta) \\
&h(\eta,\Delta) \leq \widetilde{h}^\Lambda_N(\eta;\Delta+\Lambda) + \Sigma+2\beta\left(\Delta+\Lambda\right)\;.
\label{eq:histogram}
\end{split}
\end{equation}
The derivation of Eq.~\eqref{eq:histogram} can be found in Appendix~\ref{app:hist_err}, the truncation errors $\beta^{(G,L)}$ for the Gaussian and Lorentzian are given in Eqs.~\eqref{eq:trunc_err_lor}, \eqref{eq:trunc_err_method2} in Appendix~\ref{app:truncation_err}.
Lastly, the tails $\Sigma$ (see the definition in Eq.~\eqref{eq:kernel_definition}) for the Lorentz and Gaussian kernels are bounded by
\begin{equation}
    \Sigma^{(L)} \leq \frac{2\lambda}{\pi \Lambda}\,, \ \ \ \  \Sigma^{(G)} \leq \exp\big( -\frac{\Lambda^2}{2\lambda^2}  \big)\,.
    \label{eq:sigmaG}
\end{equation}
%

%
%
%
%
%
%

\begin{figure*}[thb]
    \includegraphics[width=0.32\textwidth]{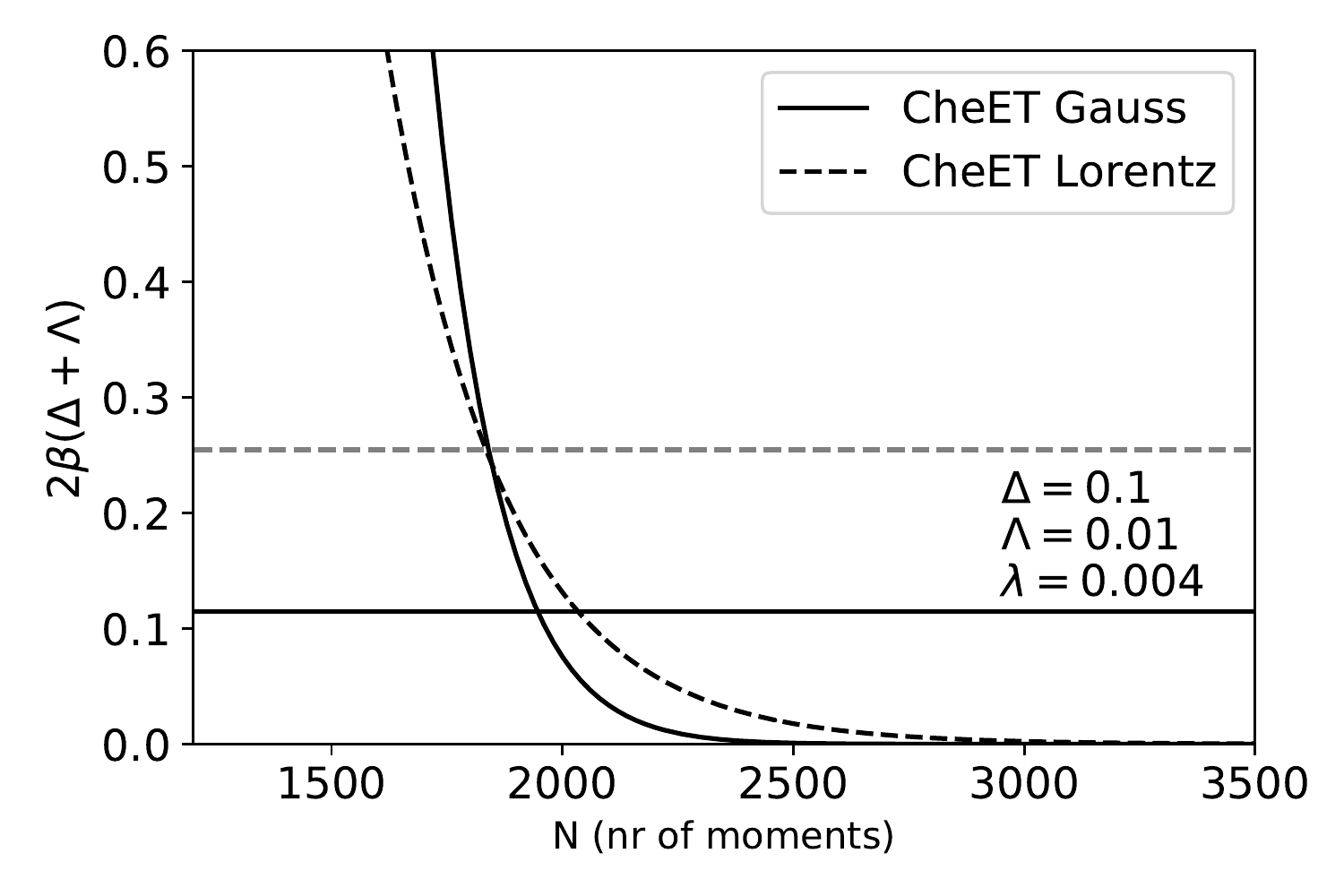}  
  \includegraphics[width=0.32\textwidth]{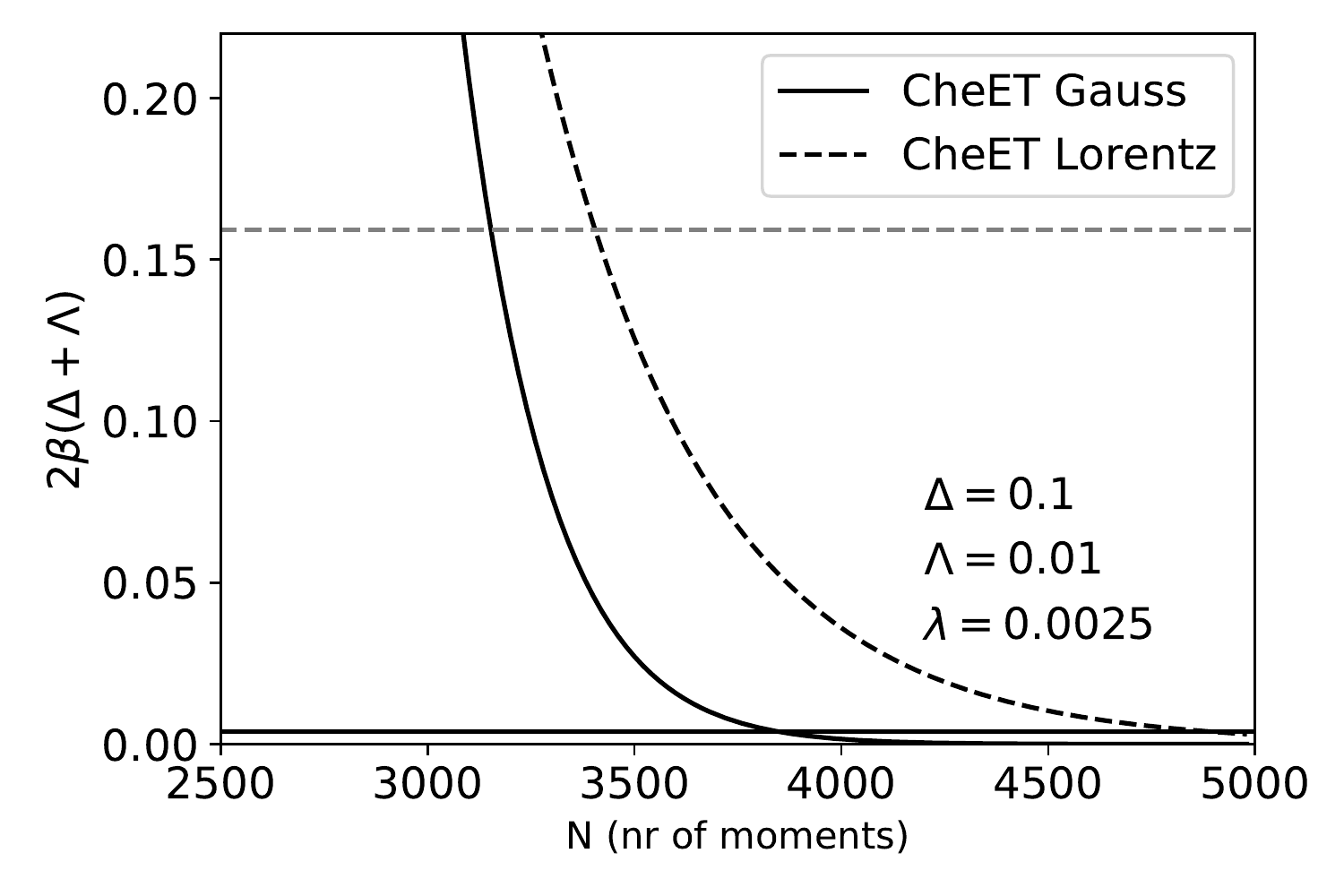}
\includegraphics[width=0.32\textwidth]{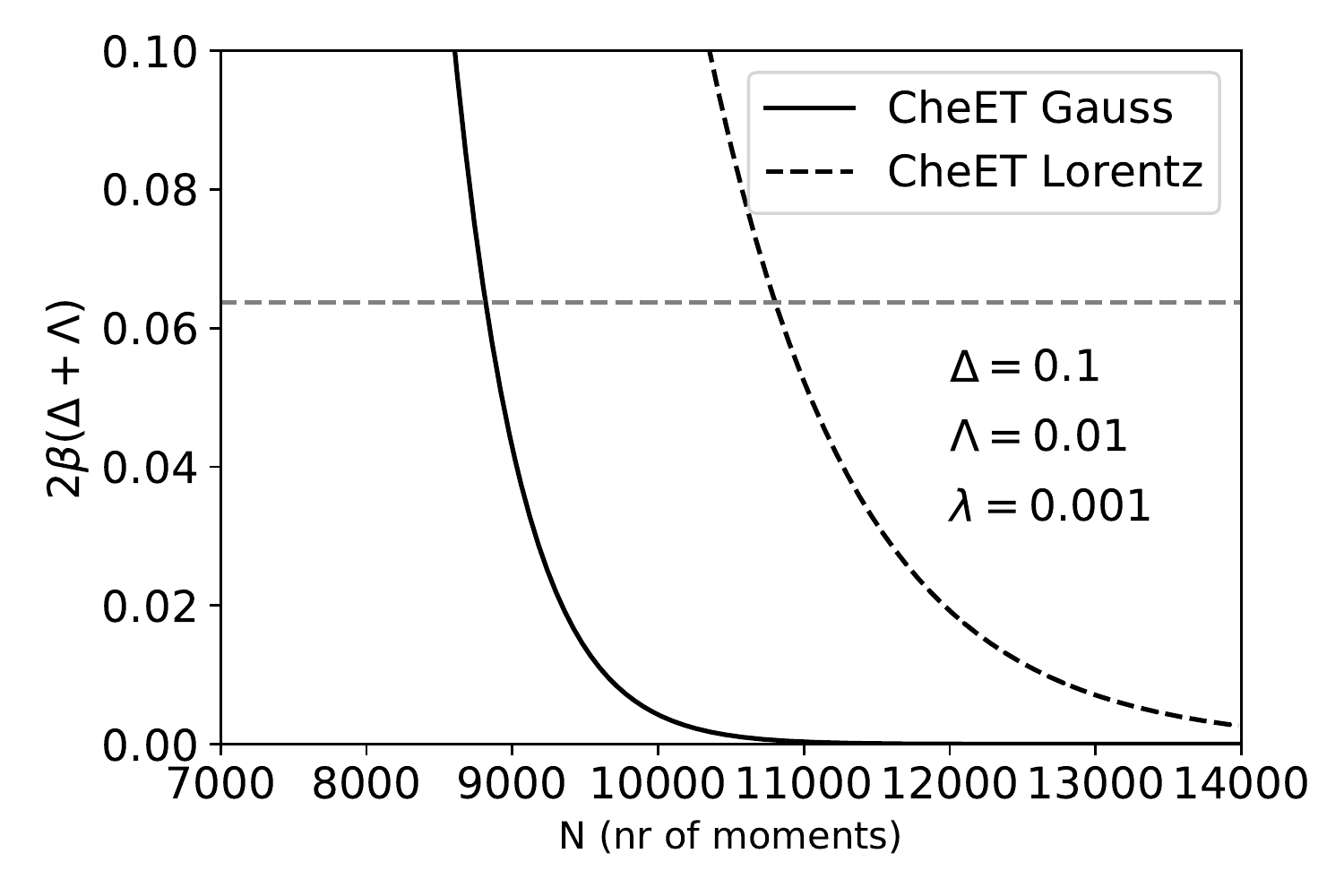}

  \caption{Truncation error as a function of number of moments, for a histogram bin $2\Delta=0.2$. Kernels resolution is $\Lambda=0.01$, using $\lambda=0.004,\ 0.0025\,, 0.001$ are shown respectively on the left, central and right panels. The dashed curves correspond to the Lorentzian kernel. The horizontal dashed and solid lines in each plot corresponds to $\Sigma^{(L)}$ and $\Sigma^{(G)}$ (the tail bound). In the right panel $\Sigma^{(G)}$ is already orders of magnitude smaller and not visible at this scale.}
  \label{fig:truncation-error}
\end{figure*}

\subsection{Case study}

We will consider an example of the signal reconstruction of both a discrete and continuous spectrum, in terms of a histogram. To generate the synthetic data we will use a function
\begin{equation}
g(x) = \sqrt{x-0.8} \exp\big[ -\frac{x}{0.35} \big]
\label{eq:fake_signal}     
\end{equation}
which is qualitatively similar to nuclear responses in the quasi-elastic regime.

\begin{itemize}
\item {\it Discrete signal}

In this case we generate the synthetic data by mimicking a many-body calculation for which the spectrum has a discrete form as in Eq.~\eqref{eq:spectral_density_expansion}. 
For this study, the spectrum is generated as a uniform random distribution of 500 delta peaks  with strengths taken from Eq.~\eqref{eq:fake_signal} (the total strength is normalized to 1) in a range $(-0.8,1)$.

\item {\it Continuous signal}

We generated a continuous function using directly Eq.~\eqref{eq:fake_signal}, with the total strength normalized to 1.
\end{itemize}

In our example we take the width of histogram bins to be $2\Delta= 0.2$.
 For the CheET method we set a desired kernel resolution to be $\Lambda=0.01$. This is driven by the following observation. Looking at Eq.~\eqref{eq:histogram}, we see that when the kernel is accurate enough ($\Sigma$ is small) and we keep the truncation error $\beta$ low (i.e. we use a sufficient number of moments), the uncertainty is driven by $(\tilde{h}_{\Delta+\Lambda}-\tilde{h}_{\Delta-\Lambda})$. Therefore we expect the error to be roughly proportional to $\Lambda/\Delta$. For the chosen values of $\Delta$ and $\Lambda$, we keep it at the order of $ \approx 10\%$.

 Having set $\Lambda$, we should choose $\lambda$, so that the tails of the distributions $\Sigma$ (see Eq.~\eqref{eq:sigmaG}) are small enough. Finally, knowing $\lambda$, the truncation error $\beta$ as a function of number of moments $N$ can be estimated. 
 In Fig.~\ref{fig:truncation-error}, we show the total truncation error $2\beta(\Delta+\Lambda)$ and the tail bound $\Sigma$ in relation to $(\lambda,N)$ for a chosen $\Delta=0.1$ and $\Lambda=0.01$, both for the Gaussian and Lorentz kernels. The horizontal lines correspond to $\Sigma^{(L,G)}$. In the right panel where $\lambda=0.001$, $\Sigma^{(G)}$ is already negligibly small and not visible. 
A compromise between the number of used moments $N$ and the desired accuracy $\lambda$ has to be found.
From the central panel of Fig.~\ref{fig:truncation-error}, we conclude that $\lambda=0.0025$ is good enough to quench $\Sigma^{(G)}$, while keeping the number of moments $N=4000$. For this value of $N$, and for the Lorentzian kernel the truncation error is smaller than $\Sigma^{(L)}$.

\begin{figure}[hbt]
    \includegraphics[width=0.5\textwidth]{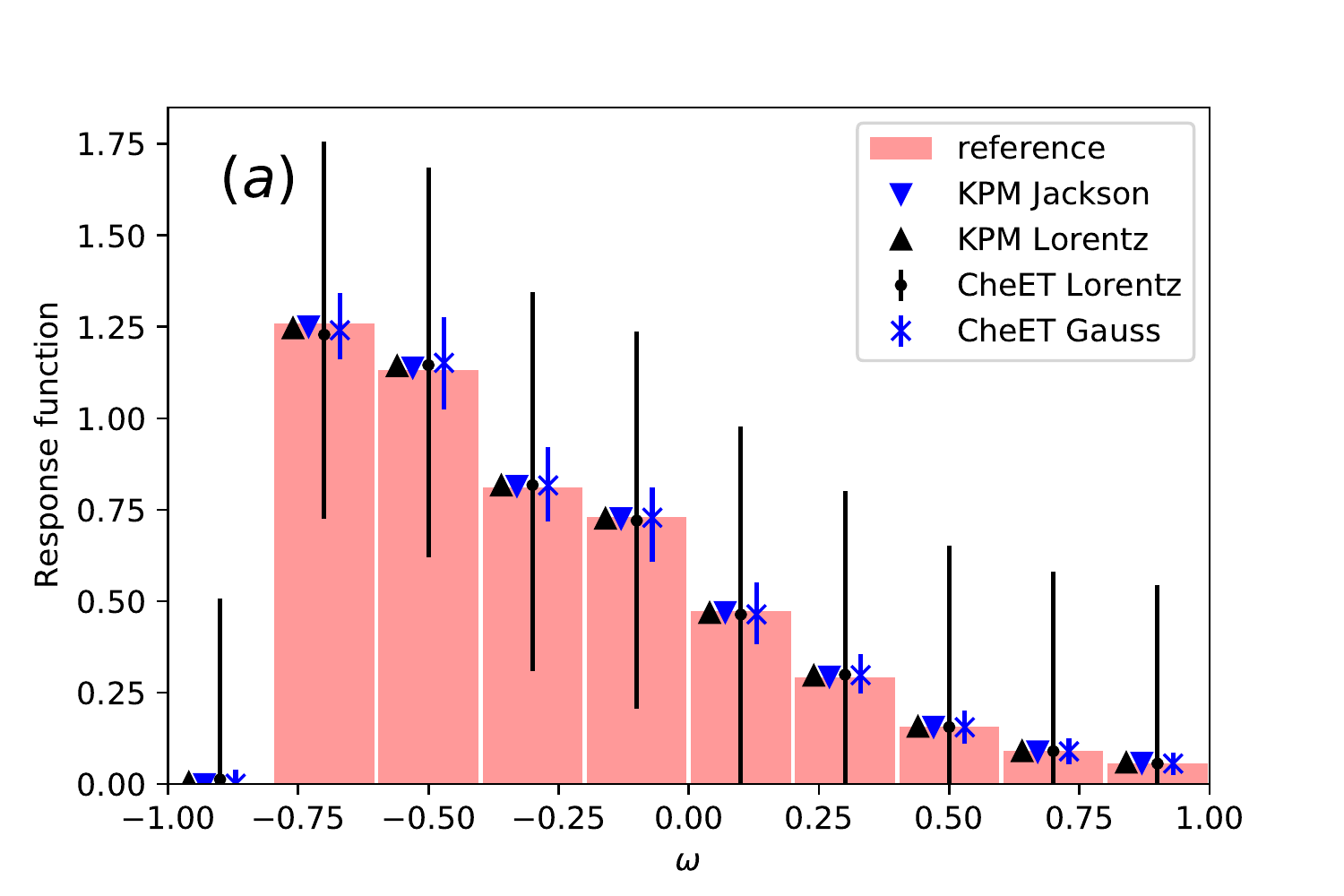}  
  \includegraphics[width=0.5\textwidth]{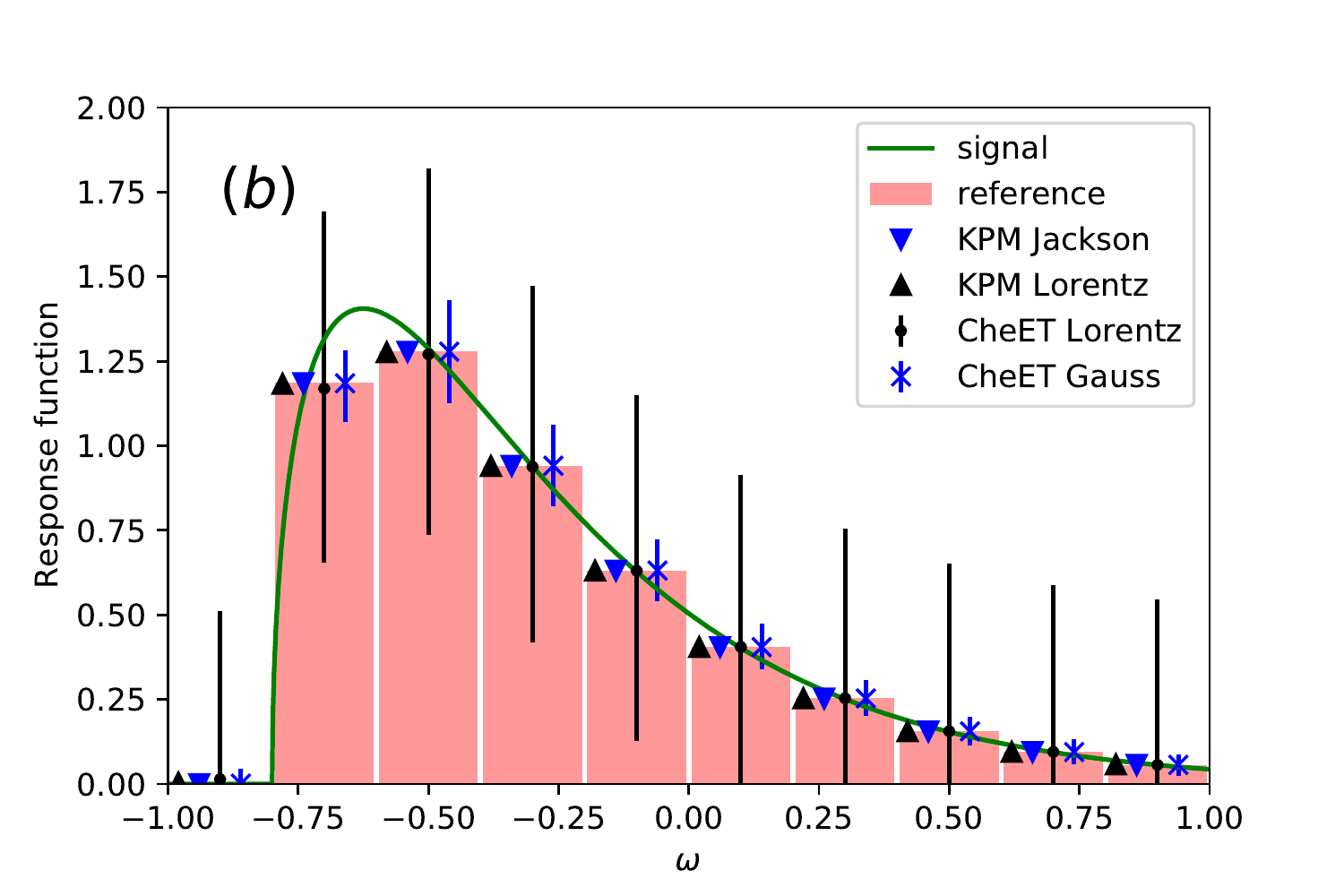}

  \caption{Signal reconstruction for a discrete case (upper panel) and a continuous case (lower panel). In both cases, the CheET Lorentz errorbars were rescaled by factor 0.5. }
  \label{fig:histograms}
\end{figure}

The results for both discrete and continuous signals for this setup are shown in Fig.~\ref{fig:histograms}. All the results, both KPM and CheET, correspond to $\widetilde{h}^\Lambda_N(\eta;\Delta)$, i.e. truncated expansion of Eq.~\eqref{eq:hist_bin_trans} with the same number of moments $N=4000$. All the four predictions give similar results which stay in agreement with the reference signal. However, the error estimation, given in Eq.~\eqref{eq:histogram}, is not available for the KPM approach. The large errors for CheET Lorentz come mostly from $\Sigma^{(L)} \propto \lambda/\Lambda$. They have been divided by factor 2 in Fig.~\ref{fig:histograms}, to fit them in the plots. They could be diminished by improving the precision $\lambda$ which would consequently require a larger number of moments. From the right panel of Fig.~\ref{fig:truncation-error} we see that diminishing this error by factor $2.5$ would require $\lambda=0.001$ and so over $N=10000$ moments.
At the same time, the CheET Gaussian gives much better uncertainty estimation. The tail bound $\Sigma^{(G)}$ and truncation error $\beta$ are negligible in this case, and the uncertainty is driven by the difference $\tilde{h}^\Lambda(\eta, \Delta+\Lambda)-\tilde{h}^\Lambda(\eta, \Delta-\Lambda)$. 

\section{Summary and Conclusions}
\label{sec:summary}

Predicting the dynamical response of strongly coupled many-body systems is a problem of central importance in nuclear physics since most of the experimental information comes from scattering cross sections. A quantitative understanding of many-body dynamics is also crucial in cold atoms experiments and quantum chemistry. In the linear response regime, the scattering cross section is related to the local spectral density, a notoriously difficult observable to evaluate in ab-initio methods. In this work we have presented a method for the reconstruction of the spectral density starting from an expansion in terms of Chebyshev polynomials using earlier results discussed in the context of quantum algorithms~\cite{roggero2020C}. This idea is similar in spirit to both the Kernel Polynomial Method (popular in condensed matter) and to the Lorentz Integral Transform method (employed in nuclear physics). Importantly, the approach presented here allows for a systematic control of the errors in the reconstruction, a key ingredient which is in general not easily achievable in both of the above mentioned techniques. 

Our results are an important step which will directly allow us to perform a full ab-initio calculation of dynamical response functions in many-body systems. In particular, they pave the way to extend the LIT-CC calculations~\cite{Bacca2013,Bacca2014,sobczyk2021ab} to compute observables for which the inversion procedure may be numerically unstable. The approach presented in this work can also be beneficial as an extension of KPM in applications using Tensor Networks~\cite{Yang_2020,Papaefstathiou_2021} and the new error bounds on the histogram discretization will provide additional guidance for the design of quantum algorithms for the estimation of the spectral density~\cite{roggero2020C,Rall2020},

In the future work we plan to address two further issues. First of all, the error estimates derived here are not necessarily tight (especially the truncation error for the Gaussian kernel in Appendix~\ref{app:git_trunc}) and it will be beneficial to improve the accuracy of the bounds. Secondly, the present method does not allow to estimate another major source of systematic bias in these calculations: the presence of an artificially discrete spectrum coming from the need to carry the many-body simulation in a finite basis. This is taken care of in the LIT framework by a careful choice of the energy resolution of the kernel. Performing a benchmark of these strategies in solvable models will be an important step forward that we will address in the future.


\begin{acknowledgements}
We thank S. Bacca and G. Hagen for useful discussions. 
J.E.S. acknowledges the support of the Humboldt Foundation through a Humboldt Research Fellowship for Postdoctoral Researchers. 
This work was supported in part by the U.S. Department of Energy, Office of Science, Office of Nuclear Physics, Inqubator for Quantum Simulation (IQuS) under Award Number DOE (NP) Award DE-SC0020970 and the Deutsche
Forschungsgemeinschaft (DFG)
through the Cluster of Excellence ``Precision Physics, Fundamental
Interactions, and Structure of Matter" (PRISMA$^+$ EXC 2118/1) funded by the
DFG within the German Excellence Strategy (Project ID 39083149)
\end{acknowledgements}

\bibliography{biblio}

%
%
%
%
%
%
\appendix
%
%
%
%
%
%
\section{Chebyshev moments for integral kernels}
\label{app_A}
In this Appedix we provide a complete derivation of the Chebyshev moments $c_k$ for the Lorentzian and Gaussian kernels defined in Eqs.~\eqref{eq:lorentzianKer} and ~\eqref{eq:gaussianKer},
\begin{equation}
K^{(G,L)}(\nu,\omega;\lambda) = \sum_{k=0}^\infty c_k(\nu,\lambda) T_k(\omega)\;,
\label{eq:kernel_exp}
\end{equation}
used in the main text.

\subsection{Moments for Lorentzian kernel}
\label{subapp_A1}

Following Ref.~\cite{jp040356n} the Chebyshev expansion of the Lorentzian can be written as 
\begin{equation}
\begin{split}
\label{eq:lor_exp}
K^{(L)}(\nu,\omega;\lambda) &= \frac{1}{\pi}\sum_{k=0}^\infty (2-\delta_{k,0})\mathcal{R}\left[ D_\lambda(\nu)Z_\lambda(\nu)^k\right] T_k(\omega)\\
\end{split}
\end{equation}
where $\mathcal{R}[z]$ is the real part of $z$ and the two functions are defined as
\begin{equation}
\begin{split}
&D_\lambda(\nu)^{-1} = \sqrt{1-(\nu+i\lambda)^2}\\
&Z_\lambda(\nu) = (\nu+i\lambda) - i D_\lambda(\nu)^{-1}\;.
\end{split}
\end{equation}

If we consider the situation where $(\nu^2+\lambda^2)<1$ we can express the second factor explicitly as
\begin{equation}
\begin{split}
Z_\lambda(\nu) &= -i \exp\left(i\arcsin(\nu+i\lambda)\right)=-i e^{-\lambda}e^{i\nu}\;.
\end{split}
\end{equation}
If we also decompose the first factor in polar coordinates $D_\lambda(\nu)^{-1}=\rho e^{i\theta}$ we can write compactly the coefficient in the Chebyshev expansion as
\begin{equation}
\begin{split}
\label{eq:lorentz_coeffs}
c_k^{(L)}(\nu;\lambda) &= 
\mathcal{R}\left[D_\lambda(\nu)Z_\lambda(\nu)^k\right] \\
&= \rho^{-1}e^{-k\lambda}\cos\left(k\left(\nu-\frac{\pi}{2}\right)-\theta\right)\;.
\end{split}
\end{equation}

\subsection{Moments for Gaussian kernel}

We start the discussion by first recalling the Chebyshev expansion of a Gaussian function
\begin{equation}
\frac{1}{\sqrt{2\pi}\lambda}e^{-\frac{\omega^2}{2\lambda^2}} = \sum_{k=0}^\infty a_k\left(\lambda\right) T_k(\omega)\;,
\end{equation}
where the moments $a_k(\lambda)$ are given explicitly as
\begin{align}
 a_n =
   \begin{dcases*}
    \frac{\gamma_n}{\sqrt{2\pi}\lambda} i^{\frac{n}{2}} \exp\left( -\frac{1}{4\lambda^2}\right) J_{n/2}\left(\frac{i}{4\lambda^2}\right) & for even $n$\\
     0 & for odd $n$
   \end{dcases*}
\end{align}
with $\gamma_n = 2-\delta_{n,0}$ and $J_n$, the Bessel function of order $n$.

Using this expansion, the Gaussian kernel can be expressed as
\begin{equation}
\label{eq:gauss_exp_base}
K^{(G)}\left(\nu, \omega; \lambda\right) =\sum_{k=0}^\infty a_k\left(\frac{\lambda}{2}\right) T_k\left(\frac{\nu-\omega}{2}\right)\;,
\end{equation}
in terms of Chebyshev polynomials depending on both variables $\nu$ and $\omega$. Our goal is instead to find a decomposition in terms of polynomials in $\omega$ of the form of Eq.~\eqref{eq:kernel_exp}

The procedure proposed in Ref.~\cite{roggero2020C} to obtain the moments $c_k(\nu,\lambda)$ proceeds as follows: we first perform the expansion
\begin{equation}
\label{eq:bivariate_expansion}
T_k\left(\frac{\nu-\omega}{2}\right) = \sum_{m=0}^\infty b_{m}^k(\nu) T_m(\omega)\;,
\end{equation}
with expansion coefficients given by
\begin{equation}
b_m^k(\nu) = \frac{\gamma_m}{\pi}\int_{-1}^1 \frac{d\omega}{\sqrt{1-\omega^2}}T_k\left(\frac{\nu-\omega}{2}\right) T_m(\omega)\, .
\end{equation}
Apart from the weight factor $1/\sqrt{1-\omega^2}$, the integrand is a polynomial in $\omega$ of maximum degree $D=k+m$. This in turn implies that, if we can perform the integration exactly using Gauss-Chebyshev quadrature as
\begin{equation}
\label{eq:gauss_cheb_int}
b_m^k(\nu) = \frac{\gamma_m}{\pi} \sum_{i=1}^L w_i T_k\left(\frac{\nu-\omega_i}{2}\right) T_m(\omega_i)
\end{equation}
with weights $w_i=\frac{\pi}{L}$ and Chebyshev nodes
\begin{equation}
\omega_i = \cos\left(\pi\frac{2i-1}{2L}\right)\;,
\end{equation}
provided we choose $L>(D+1)/2$. In the following we will take $L=L_{m,k}=\lceil(m+k+1)/2\rceil$.

Now, by realizing that, for any choice of $\nu$, the function $T_k((\nu-\omega)/2)$ is a polynomial of order $k$ in $\omega$, the sum in Eq.~\eqref{eq:bivariate_expansion} can be truncated at order $m\geq k$ without incurring in an approximation error. This implies that we can take, for a given $k$, the truncation in Eq.~\eqref{eq:gauss_cheb_int} as
\begin{equation}
L = L_{k,k} = \left\lceil\frac{2k+1}{2}\right\rceil = k+1\;.
\end{equation}

Suppose now that we approximate the Gaussian kernel with a truncated sum of the form
\begin{equation}
\begin{split}
K^{(G),N}(\nu,\omega;\lambda) &= \sum_{k=0}^N a_k\left(\frac{\lambda}{2}\right) T_k\left(\frac{\nu-\omega}{2}\right)\\
& = \sum_{k=0}^N \sum_{m=0}^\infty a_k\left(\frac{\lambda}{2}\right)  b_{m}^k(\nu) T_m(\omega)\;,
\end{split}
\end{equation}
with a corresponding truncation error $\beta^{(G)}_N$ derived in Ref.~\cite{roggero2020C} and discussed in more detail in the next Appendix.
In the original derivation in Ref.~\cite{roggero2020C} the expansion in $m$ was truncated at $m=N\geq k$ resulting in
\begin{equation}
\label{eq:nested_gaussian_exp}
\begin{split}
K^{(G),N}(\nu,\omega;\lambda) & = \sum_{k=0}^N \sum_{m=0}^N a_k\left(\frac{\lambda}{2}\right)  b_{m}^k(\nu) T_m(\omega)\\
&=\sum_{m=0}^N \left(\sum_{k=0}^N  a_k\left(\frac{\lambda}{2}\right)  b_{m}^k(\nu)\right) T_m(\omega)\\
&=\sum_{m=0}^N \widetilde{c_m}^{[N]}(\nu,\lambda) T_m(\omega)\;,
\end{split}
\end{equation}
with expansion coefficients given explicitly as
\begin{equation}
\begin{split}
\widetilde{c_m}^{[N]}(\nu,\lambda) = \sum_{k=0}^N\sum_{i=1}^{L_{k,k}}\frac{\gamma_m}{L_{k,k}}a_k\left(\frac{\lambda}{2}\right)T_k\left(\frac{\nu-\omega^k_i}{2}\right) T_m(\omega^k_i)\;,
\end{split}
\end{equation}
with Chebyshev nodes $\omega^k_i$ depending explicitly on $k$ due to the corresponding $k$ dependence of the number of terms $L_{k,k}$.
As in Ref.~\cite{roggero2020C} this can be removed by performing a further simplification by choosing $L_{k,k}=N+1$ independent on $k$. As the discussion on the Gaussian quadrature formula provided above shows, this does not introduce further errors and results in a modest $\mathcal{O}(N)$ increase in number of summands. The final expression for the expansion coefficients is then
\footnote{We note two typos in Ref.~\cite{roggero2020C} with $(N+1)$ being indicated as $N$ and the numerator in Eq.~\eqref{eq:kindep_nodes} being quoted to be (effectively) $2i+1$ instead of $2i-1$. These do not affect any of the results discussed there but are important for a correct implementation of the expansion coefficients.}
\begin{equation}
\label{eq:c_w_kindep_nodes}
\widetilde{c_m}^{[N]}(\nu,\lambda) =\frac{\gamma_m}{N+1} \sum_{k=0}^N\sum_{i=1}^{N+1}a_k\left(\frac{\lambda}{2}\right)T_k\left(\frac{\nu-\widetilde{\omega_i}}{2}\right) T_m(\widetilde{\omega_i})\;,
\end{equation}
where we have defined the $k$-independent nodes
\begin{equation}
\label{eq:kindep_nodes}
\widetilde{\omega}_i = \cos\left(\pi\frac{2i-1}{2(N+1)}\right)\;.
\end{equation}

We will call the scheme presented so far {\it method 1}. It has one main disadvantage with respect to the construction for the Lorentz kernel above: evaluation of the kernel (or equivalently the integral transform) at different frequencies $\nu$ incurs in a cubic cost with the number of terms $N$ whereas for the Lorentz kernel this cost is only linear in $N$. Another drawback of the present construction is that the coefficients $\widetilde{c_m}(\nu,\lambda)$ in the kernel expansion of Eq.~\eqref{eq:nested_gaussian_exp} and provided explicitly in Eq.~\eqref{eq:c_w_kindep_nodes} are not the same as the $c_k(\nu,\lambda)$ coefficients obtained from a direct uni-variate expansion of the kernel in the $\omega$ frequency as in Eq.~\eqref{eq:kernel_exp}. As we will see below this might result in a worse truncation error, at fixed $N$, than one could obtain if the latter expansion coefficients were known analytically (as in the case of the Lorentzian).

\begin{figure}[hbt]
    \includegraphics[width=0.5\textwidth]{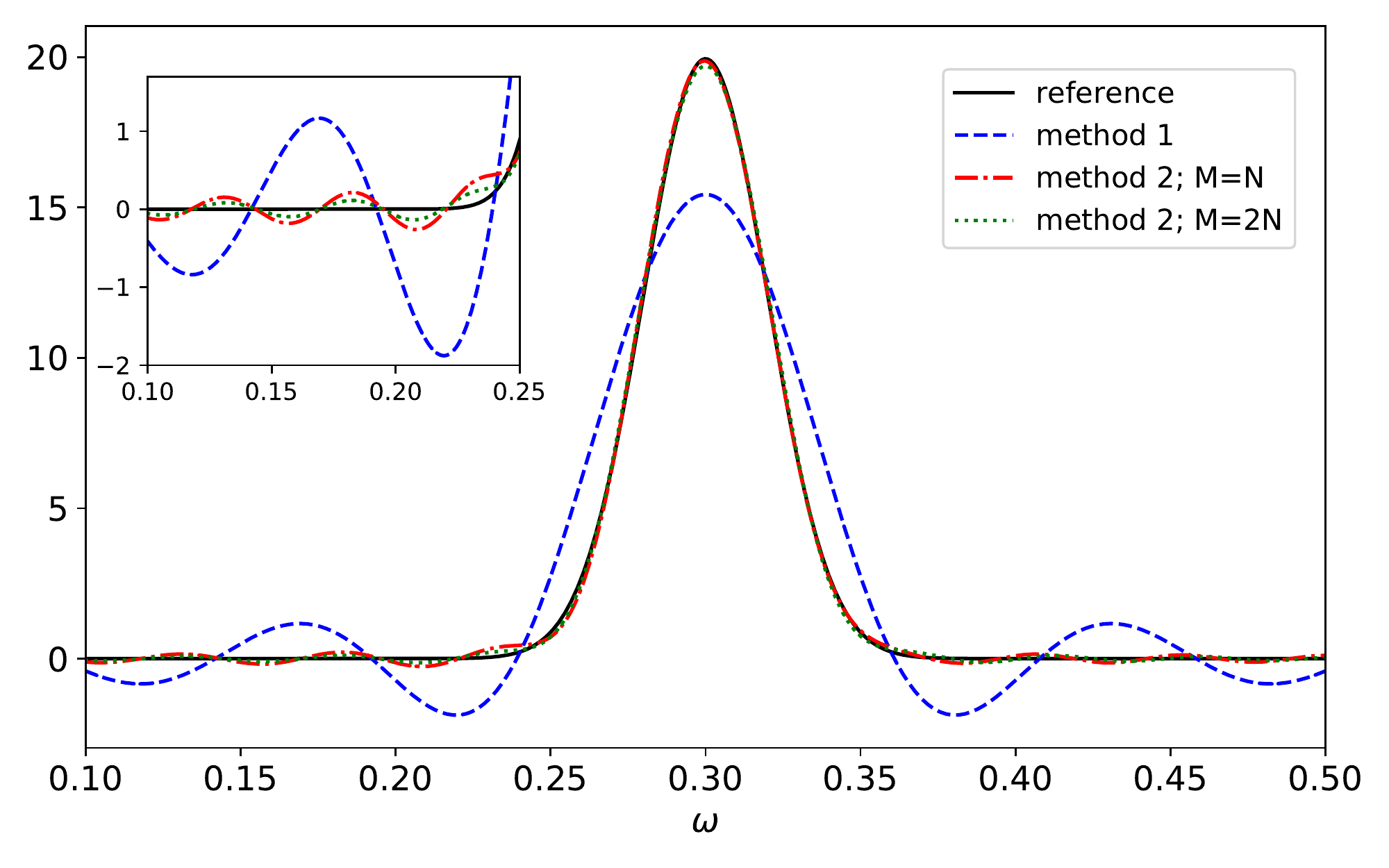} 

  \caption{Reconstruction of the Gaussian signal ($\lambda=0.05$, $\nu=0.3$) using two different sets of expansion coefficients given by Eqs.~\eqref{eq:c_w_kindep_nodes} ({\it method 1}) and \eqref{eq:gauss_method2} ({\it method 2}). In all cases we use $N=120$ moments. For method 2 we show both $M=120$ and $M=240$ integration points. }
  \label{fig:gauss_exp}
\end{figure}

To address both of these problems, here we also consider a second approach that directly estimates the ``exact'' coefficients
\begin{equation}
c_k(\nu,\lambda)= \frac{\gamma_k}{\pi}\int_{-1}^1 \frac{d\omega}{\sqrt{1-\omega^2}}K^{(G)}\left(\nu, \omega; \lambda\right) T_k(\omega)\, .
\label{eq:ck}
\end{equation}
by approximating the integral with a Gauss-Chebyshev quadrature using a large number of nodes $M> N$ for a target truncation level $N$
\begin{equation}
\begin{split}
c_k^{[N,M]}(\nu,\lambda)= \sum_{m=1}^{M} \frac{\gamma_k}{M} K^{(G)}(\nu, \omega_m; \lambda) T_k(\omega_m)
\label{eq:gauss_method2}
\end{split}
\end{equation}
with $\omega_m = \cos (\pi\frac{2m-1}{2M})$ the Chebyshev nodes as above. This construction, {\it method 2} has the main advantage of resulting in a faster evaluation of the kernel and integral transform. The coefficients $c_k^{[N,M]}(\nu,\lambda)$ converge close to the exact ones $c_k^{(G)}(\nu,\lambda)$ in the large $M$ limit and this appears to considerably reduce the truncation error. For the results presented in the main text we use {\it method 2}, i.e. coefficients of Eq.~\eqref{eq:gauss_method2}. To support our choice, in Fig.~\ref{fig:gauss_exp} we present a simple comparison between the two expansions, which shows a faster conversion of {\it method 2}.

\section{Bounds on the truncation error}
\label{app:truncation_err}
In this section we provide the proofs for the bounds on the truncation error $\beta$ from Eq.~\eqref{eq:beta} of the main text.

\subsection{Lorentz Kernel}
Following the derivation in Appendix~\ref{subapp_A1}, for the case of the Lorentzian we have directly a closed-form expression for the expansion coefficients of the integral transform. We can then estimate $\beta$ by first using 
\begin{equation}
\begin{split}
\delta^{(L)}_{trunc}&=\sup_{\nu\in[-1,1]} \left|\sum_{k=N+1}^\infty c_k(\nu;\lambda)m_k\right|\\
&\leq \sup_{\nu\in[-1,1]} \sum_{k=N+1}^\infty \left| c_k(\nu; \lambda)\right|
\end{split}
\end{equation}
and then using the expression for coefficients $c_k$ from Eq.~\eqref{eq:lorentz_coeffs}:
\begin{equation}
\begin{split}
\delta^{(L)}_{trunc}&\leq\sup_{\nu\in[-1,1]}  \frac{2}{\pi\rho}\sum_{n=N+1}^\infty e^{-\lambda n}\left|\cos\left(n\left(\nu-\frac{\pi}{2}\right)-\theta\right)\right|\\
&\leq\frac{2}{\pi\rho}\sum_{n=N+1}^\infty e^{-\lambda n}\\
&\leq\frac{2}{\pi\rho}\int_{N}^\infty dx e^{-\lambda x}=\frac{2e^{-\lambda N}}{\lambda\rho\pi}\;.
\end{split}
\end{equation}
Note at this point that we have also
\begin{equation}
\rho=((1+\nu^2+\lambda^2)^2-4\nu^2)^{1/4}\geq \sqrt{1-\nu^2}\;,
\end{equation}
but this lower bound is not useful if we keep the spectrum in $[-1,1]$ as it approaches zero. One option is to rescale the Hamiltonian operator in order to work in a smaller interval, an alternative is to use instead the bound
\begin{equation}
\begin{split}
\rho &\geq ((2+\lambda^2)^2-4)^{1/4} = \sqrt{\lambda}(4+\lambda^2)^{1/4}\geq \sqrt{2\lambda}
\end{split}
\end{equation}
Using this we have the error bound used in the main text
\begin{equation}
\label{eq:trunc_err_lor}
\beta^{(L)}_N = \frac{\sqrt{2}e^{-\lambda N}}{\lambda^{3/2}\pi}\;.
\end{equation}

\subsection{Gaussian kernel Kernel}
\label{app:git_trunc}

We first compute the bound for the {\it method 1} approximation in Eq.~\eqref{eq:nested_gaussian_exp}. First note that we can directly bound the error in the integral tansform in terms of the error in the kernel using
\begin{equation}
\begin{split}
|\Phi(\nu;\lambda)-&\Phi_N(\nu;\lambda)| \\
&=\left|\int\!\!\! d\omega S(\omega)\!\left(K(\nu,\omega;\lambda)-K_N(\nu,\omega;\lambda)\right)\right|\\
&\leq\int d\omega S(\omega)\left|K(\nu,\omega;\lambda)-K_N(\nu,\omega;\lambda)\right|\\
&\leq \sup_{\omega\in[-1,1]}\left|K(\nu,\omega;\lambda)-K_N(\nu,\omega;\lambda)\right|
\end{split}
\end{equation}
where we used that $S(\omega)\geq0$ and normalized to one. Since we perform the exact expansion of the two-variable Chebyshev polynomial $T_k((\nu-\omega)/2)$ using Eq.~\eqref{eq:bivariate_expansion} truncated at $m=N\geq k$, we find directly that
\begin{equation}
\begin{split}
\delta_{trunc}^{(G),N} & = \sup_{\nu\in[-1,1]}\sup_{\omega\in[-1,1]}\left|K^{(G)}(\nu,\omega;\lambda)-K^{(G),N}(\nu,\omega;\lambda)\right|\\
&=\sup_{\nu\in[-1,1]}\sup_{\omega\in[-1,1]}\left|\sum_{k=N+1}^\infty a_k\left(\frac{\lambda}{2}\right)T_k\left(\frac{\nu-\omega}{2}\right)\right|\\
&\leq \sum_{k=N+1}^\infty \left|a_k\left(\frac{\lambda}{2}\right)\right|\;.
\end{split}
\end{equation}
The sum on the last line was shown in Ref.~\cite{roggero2020C} to be bounded as
\begin{equation}
\sum_{k=N+1}^\infty \left|a_k\left(\frac{\lambda}{2}\right)\right| \leq \frac{1}{2\lambda}\sqrt{\frac{\pi}{\kappa(1)}}\text{erfc}\left((N+1)\lambda\sqrt{\frac{\kappa(1)}{2}}\right)\;,
\end{equation}
where the function $\kappa$ is given by
\begin{equation}
\label{eq:kappa}
\kappa(x)=\frac{\log(x+\sqrt{1+x^2})}{2}-\frac{1}{4x}\frac{\left(x-1+\sqrt{1+x^2}\right)^2}{x+\sqrt{1+x^2}}\;.
\end{equation}
Therefore for {\it method 1} the truncation error can be bounded by
\begin{equation}
\beta^{(G)}_N = \frac{1}{2\lambda}\sqrt{\frac{\pi}{\kappa(1)}}\text{erfc}\left((N+1)\lambda\sqrt{\frac{\kappa(1)}{2}}\right)\;.
\end{equation}

For the Gaussian kernel obtained with the {\it method 2} coefficients from Eq.~\eqref{eq:gauss_method2} we can find a bound on the truncation error (possibly very loose) as follows. First we can use the expansion in Eq.~\eqref{eq:gauss_exp_base} to re-express the new expansion coefficients as
\begin{equation}
\begin{split}
c_k^{[N,M]}(\nu,\lambda)&= \sum_{m=1}^{M} \frac{\gamma_k}{M}\sum_{n=0}^\infty a_n\left(\frac{\lambda}{2}\right) T_n\left(\frac{\nu-\omega_m}{2}\right) T_k(\omega_m)\;.
\end{split}
\end{equation}
The full kernel can then be written as
\begin{equation}
\begin{split}
K^{(G),N,M} = \sum_{k=0}^N\sum_{m=1}^{M} & \frac{\gamma_k}{M}\sum_{n=0}^\infty  a_n\left(\frac{\lambda}{2}\right)\\
&T_n\left(\frac{\nu-\omega_m}{2}\right) T_k(\omega_m)T_k(\omega)\;.
\end{split}
\end{equation}
At this point it is convenient to define a finite order approximation to the two-variable Chebyshev coefficient as
\begin{equation}
T^{N,M}_n(\nu,\omega) = \sum_{k=0}^N \frac{\gamma_k}{M}\sum_{m=1}^{M}T_n\left(\frac{\nu-\omega_m}{2}\right) T_k(\omega_m)T_k(\omega)\;.
\end{equation}
Following the discussion used to obtain the Chebyshev coefficients from {\it method 1}, we know that
\begin{equation}
T^{N,M}_n(\nu,\omega) = T_n\left(\frac{\nu-\omega}{2}\right)\ \ \ \mathrm{for}\ N\geq n,\ \ M\geq n+1 .
\end{equation}
Using this notation we can write
\begin{equation}
K^{(G),N,M}(\nu,\omega;\lambda) = \sum_{n=0}^\infty a_n\left(\frac{\lambda}{2}\right) T^{N,M}_n(\nu,\omega)\;,
\end{equation}
while the exact kernel reads
\begin{equation}
\begin{split}
K^{(G)}(\nu,\omega;\lambda) &= \sum_{n=0}^\infty a_n\left(\frac{\lambda}{2}\right) T^{n,n+1}_n\left(\frac{\nu-\omega}{2}\right)\\
&= \sum_{n=0}^\infty a_n\left(\frac{\lambda}{2}\right) T_n\left(\frac{\nu-\omega}{2}\right)\;.
\end{split}
\end{equation}

We can then write their difference as
\begin{equation}
\begin{split}
&K^{(G)}(\nu,\omega;\lambda) - K^{(G),N,M}(\nu,\omega;\lambda) \\
&= \sum_{n=N+1}^\infty a_n\left(\frac{\lambda}{2}\right) \left(T_n\left(\frac{\nu-\omega}{2}\right)-T^{N,M}_n\left(\frac{\nu-\omega}{2}\right)\right)\;,
\end{split}
\end{equation}
provided we choose $M\geq N+1$. 

In order to bound the difference between $T_n$ and $T_n^{N,M}$ defined as
\begin{equation}
\delta_n^N(\nu,\omega) = \left|T_n\left(\frac{\nu-\omega}{2}\right)-T_n^{N,N+1}(\nu,\omega)\right|\;,
\end{equation}
we first recall that, from theorem 2.1 of~\cite{chebfun} we have, for any $\omega\in[-1,1]$, that (see also~\cite{Gunttner1980})
\begin{equation}
\delta_n^N(\nu,\omega)\leq\left(2+\frac{2}{\pi}\log(N+1)\right)\left|T_n\left(\frac{\nu-\omega}{2}\right) - p_N^*(\omega)\right|\;,
\end{equation}
where $p_N^*(\omega)$ is the optimal approximating polynomial of order at most $N$ for a given fixed choice of $\nu$ (ie. we look at $T_n((\nu-\omega)/2)$ as a function of $\omega$ only). One option is to now use Jackson's theorems~\cite{approx_book} to bound the right hand side. However, owing to the fact that $n\gg N$ for our purposes, we weren't able to obtain tight bounds in this way. The alternative used to compute the error estimates in the main text was instead to use
\begin{equation}
\left|T_n\left(\frac{\nu-\omega}{2}\right) - p_N^*(\omega)\right|\leq\left|T_n\left(\frac{\nu-\omega}{2}\right) - p_0^*(\omega)\right|\;,
\end{equation}
with $p_0^*(\omega)$ the optimal approximating constant. Using the fact that $|T_n(\omega)|\leq1$ together with Corollary 1.6.1 of~\cite{approx_book} we have $|T_n((\nu-\omega)/2) - p_0^*(\omega)|\leq1$ so that
\begin{equation}
\sup_{\nu\in[-1,1]}\sup_{\omega\in[-1,1]} \delta_n^N(\nu,\omega)\leq\left(2+\frac{2}{\pi}\log(N+1)\right)\;.
\end{equation}
In the main text we have then used the following truncation bound for {\it method 2}
\begin{equation}
\beta^{(G)}_{N,N+1} = \left(2+\frac{2}{\pi}\log(N+1)\right)\beta^{(G)}_N\;.
\label{eq:trunc_err_method2}
\end{equation}

As evident by the results in Fig.~\ref{fig:gauss_exp}, where we show a comparison between the kernel function obtained using both methods, this estimate for the truncation error of {\it method 2} is likely a very conservative upperbound and we expect in general that $\beta^{(G)}_{N,M}\leq \beta^{(G)}_N$. In future work it would be valuable to find tighter error bounds as they will impact the total error budget in the estimation of histograms of the spectral density.

%
%
%
%
%
%
\section{Error bound on histograms}
\label{app:hist_err}

We want to assess the error for $h(\eta; \Delta)$ defined in Eq.~\eqref{eq:hist_bin}.
The error has two sources, coming from the fact of using $\Sigma$-accurate kernel and from the truncation of the kernel.

Starting from the definition of a histogram of Eq.~\eqref{eq:histo_bin}, let us first notice that
\begin{equation}
\widetilde{f}^\Lambda(\eta,\eta; \Lambda) = \int_{\eta-\Lambda}^{\eta+\Lambda} d\nu K(\eta,\nu;\Lambda)\geq 1-\Sigma\;.
\end{equation}
This property also holds for larger intervals $\delta>\Lambda$ and for energies $|\omega-\eta|\leq\delta-\Lambda$ as follows
\begin{equation*}
\widetilde{f}^\Lambda(\omega,\eta; \delta) = \int_{\eta-\delta}^{\eta+\delta} d\nu K(\omega,\nu;\Lambda)\geq 1-\Sigma\;.
\end{equation*}
This is obtained by realizing that $\widetilde{f}^\Lambda(\omega,\eta; \delta)$ is at least $(1-\Sigma)$ if we can find an interval of size $2\Lambda$, centered in $\eta$ and contained in the full interval of size $2\delta$.
Since the kernel is normalized, this also implies
\begin{equation}
\begin{split}
1-\widetilde{f}^\Lambda(\omega,\eta; \delta) &= \int_{-1}^{\eta-\delta} d\nu K_\Lambda(\omega,\nu) + \int_{\eta+\delta}^{1} d\nu K_\Lambda(\omega,\nu)\\
&\leq \Sigma \quad\text{for}\quad|\omega-\eta|\geq\delta+\Lambda
\end{split}
\end{equation}
This condition allows us to construct an approximation of the window function $f$ using its transform. In fact we have the following bound
\begin{equation}\label{eq:approx_bound}
\sup_{|\omega-\eta|\in[0,\delta-\Lambda]\cup[\delta+\Lambda,\infty]} \left|f(\omega,\eta; \delta) - \widetilde{f}^\Lambda(\omega,\eta; \delta)\right| \leq \Sigma\;.
\end{equation}

The error in the two disjoint intervals has different signs, more explicitly we have
\begin{equation}
\label{eq:inner_tail}
\sup_{|\omega-\eta|\in[0,\delta-\Lambda]} \left(f(\omega,\eta; \delta) - \widetilde{f}^\Lambda(\omega,\eta; \delta)\right) \leq \Sigma\;.
\end{equation}
since the approximation is always smaller that the indicator function there, and
\begin{equation}
\label{eq:tail_bound}
\begin{split}
\sup_{|\omega-\eta|\in[\delta+\Lambda,\infty]} \left(\widetilde{f}^\Lambda(\omega,\eta; \delta)-f(\omega,\eta; \delta)\right) &=\\ \sup_{|\omega-\eta|\in[\delta+\Lambda,\infty]} \widetilde{f}^\Lambda(\omega,\eta; \delta)&\leq \Sigma\;.
\end{split}
\end{equation}

Let us also notice that
\begin{equation}
f(\omega,\eta; \delta+\Lambda) \geq \widetilde{f}^\Lambda(\omega,\eta; \delta)\quad\text{for}\quad|\omega-\eta|\leq\delta+\Lambda\;,
\label{eq:inner_val}
\end{equation}

Combining Eqs.~\eqref{eq:tail_bound},~\eqref{eq:inner_val} we find the following lower bound for any $\omega$
\begin{equation}
f(\omega,\eta; \delta+\Lambda) \geq \widetilde{f}^\Lambda(\omega,\eta; \delta) - \Sigma\;.
\end{equation}
This immediately implies the following lower bound to the histogram
\begin{equation}
h(\eta; \Delta) \geq \widetilde{h}^{\Lambda}(\eta; \Delta-\Lambda) - \Sigma\;.
\end{equation}

For the upperbound we can use instead
\begin{equation}
f(\omega,\eta; \delta-\Lambda) \leq \widetilde{f}^\Lambda(\omega,\eta; \delta)\quad\text{for}\quad|\omega-\eta|\geq\delta-\Lambda\;,
\end{equation}
together with the (inner) tail condition from Eq.~\eqref{eq:inner_tail}
\begin{equation}
\widetilde{f}^\Lambda(\omega,\eta; \delta) \geq 1-\Sigma\quad\text{for}\quad|\omega-\eta|\leq\delta-\Lambda\;.
\end{equation}
Combining these two we find the following upper bound for any $\omega$
\begin{equation}
f(\omega,\eta; \delta-\Lambda) \leq \widetilde{f}^\Lambda(\omega,\eta; \delta) + \Sigma\;.
\end{equation}
This immediately implies the following lower bound to the histogram
\begin{equation}
h(\eta; \Delta) \leq \widetilde{h}^{\Lambda}(\eta; \Delta+\Lambda) + \Sigma\;.
\end{equation}

The final result is the following two sided bound on the correct histogram
\begin{equation}
\widetilde{h}^{\Lambda}(\eta; \Delta-\Lambda) - \Sigma \leq h(\eta; \Delta) \leq \widetilde{h}^{\Lambda}(\eta; \Delta+\Lambda) + \Sigma\;.
\end{equation}

Including the truncation error, as in Eq.~\eqref{eq:histo_truncationErr}, we arrive finally at Eq.~\eqref{eq:histogram}.

\end{document}